\documentclass[12pt,aps,amsmath,latexsym,amsfonts]{JHEP3}

\usepackage{epsfig}
\usepackage{amsfonts,amssymb,amsmath}
\usepackage[hang]{subfigure}

\newcommand{\be}{\begin{equation}}
\newcommand{\ee}{\end{equation}}
\newcommand{\bea}{\begin{eqnarray}}
\newcommand{\eea}{\end{eqnarray}}
\newcommand{\beal}{\begin{aligned}}
\newcommand{\eeal}{\end{aligned}}

\newcommand{\phistar}{\ensuremath{\phi_{\star}}}
\newcommand{\varphistar}{\ensuremath{\varphi_{\star}}}
\newcommand{\rhostar}{\ensuremath{\rho_{\star}}}
\newcommand{\mstar}{\ensuremath{m_{\star}}}
\newcommand{\Veff}{\ensuremath{V_{\text{eff}}}}

\title{Astrophysical black holes in screened modified gravity}

\author{Anne-Christine Davis$^1$\thanks{Email: acd@damtp.cam.ac.uk},
Ruth Gregory$^{2,3}$\thanks{Email: r.a.w.gregory@durham.ac.uk},
Rahul Jha$^1$\thanks{Email: r.jha@damtp.cam.ac.uk},
Jessica Muir$^1$\thanks{Now at University of Michigan: jlmuir@umich.edu}\\
$^1${\it Department of Applied Mathematics and Theoretical Physics,
Centre for Mathematical Sciences, University of Cambridge, 
Wilberforce Road, Cambridge, CB3 0WA, U.K.}\\
$^2${\it Centre for Particle Theory, South Road, Durham, DH1 3LE, UK}\\
$^3${\it Perimeter Institute, 31 Caroline Street North, Waterloo, ON, N2L 2Y5,
Canada}
}

\abstract{
Chameleon, environmentally dependent dilaton, and symmetron gravity are 
three models of modified gravity in which the effects of the additional scalar 
degree of freedom are screened in dense environments. They have been 
extensively studied in laboratory, cosmological, and astrophysical 
contexts. In this paper, we  present a preliminary investigation into whether 
additional constraints can be provided by studying these scalar fields around 
black holes.  By looking at the properties of a static, spherically symmetric 
black hole,  we find that the presence of a non-uniform matter distribution
induces a non-constant scalar profile in chameleon and dilaton, but not 
necessarily symmetron gravity.  An order of magnitude estimate shows that 
the effects of these profiles on in-falling test particles will be sub-leading
compared to gravitational waves and hence observationally challenging to detect. }

\keywords{Black holes, scalar fields, no hair theorems}
\preprint{DCPT-14/03}

\begin{document}
\section{Introduction}\label{intro}

Since it was first published nearly a century ago, general relativity (GR) 
has earned its place as an incredibly successful and well verified theory of gravity
(see e.g.\ \cite{Will2005}). There are however, both theoretical and
observational reasons to consider alternatives to, and extensions of, GR. 
On the one hand, string theory provides a framework for
describing quantum gravity, but suggests that we live in more than 4 dimensions.
The consequences of these extra dimensions have not been definitively 
predicted, however, a generic feature of dimensional reduction is that
extra fields appear in the low energy gravitational sector. On the other hand,
observations of high redshift supernovae indicate that the universe 
is expanding at an accelerating rate~\cite{Riess1998,Perlmutter1999}, 
and together with microwave background~\cite{Hinshaw:2012aka,Ade:2013zuv}
and large scale structure~\cite{Abazajian:2008wr} measurements, 
suggest that around $70 \%$ of the energy density of the universe 
comes in the form of a `dark energy' -- an energy momentum density which
has a large negative pressure and is well modelled by a cosmological
constant. Since the magnitude of the cosmological constant is extremely
small by particle physics standards, explaining its stability under quantum
corrections is a challenge. Finding either a natural explanation for its measured 
value or an alternative to the constant, is a major motivation for 
developing and studying modified theories of gravity (see \cite{Clifton2011}
for a review of various approaches). Additionally, studying modified 
gravity theories allows us to better explore where GR has been tested 
rigorously and to constrain the low energy properties of quantum gravity theories.

Scalar-tensor theories, which modify gravity by introducing new, 
non-minimally coupled scalar fields are an extensively studied alternative to GR.  
They are theoretically attractive because such light scalar fields are generically
predicted in the low energy limit of string theory. For a scalar field to affect 
cosmological expansion, its mass must be of the order of the Hubble 
scale, $H_0\sim 10^{-33}\,\text{eV}$.  If it interacts with matter however, 
the presence of a light scalar field will result in a long-range fifth force 
and would thus be subject to tight constraints from laboratory and solar 
system tests of gravity, \cite{Will2005,Bertotti2003,Williams2004}. 
These constraints are weakened for theories in which the scalar field 
modifying gravity somehow decouples from matter, or is 
``screened'', in dense environments.    
Models with this property are appealing because they can modify the 
behaviour of gravity on large scales while recovering the behaviour 
of GR in environments where local tests have been performed. 
This paper will focus on three particular scalar-tensor theories which have been 
found to exhibit screening, known as chameleon, \cite{Khoury2004}, 
environmentally dependent dilaton, \cite{Brax2010}, and 
symmetron, \cite{Hinterbichler2011}, modified gravity.

These models have been studied and constrained using laboratory
\cite{Mota2008,Upadhye2012b,Upadhye2012,Upadhye2012a}, 
solar system~\cite{Bertotti2003}, cosmological~\cite{Jain2010}, and 
astrophysical~\cite{Jain2012} tests. These investigations all have 
one thing in common: they probe gravity in a regime where gravitational 
fields and space-time curvatures are relatively weak.  In the near 
future, the direct detection of gravitational waves from compact binary 
systems will allow us to constrain the behaviour of gravity in the strong 
field, large curvature regime. Accordingly, attention has increasingly 
focussed on efforts to test gravity by studying the dynamics of compact 
objects such as neutron stars and black holes, \cite{Yunes2013,Psaltis2008}.  
It is thus natural to ask whether observations of black holes 
might provide new constraints on screened modified gravity.

Studying black holes in the context of a modified 
gravity theories inevitably leads to the re-examination of the 
uniqueness of exact solutions, 
i.e. ,whether the extra physical fields add extra degrees of freedom to 
black hole solutions, usually referred to as ``black hole hair'', 
\cite{Ruffini}. A slightly looser definition of ``hair'', in terms of 
non-trivial scalar profiles rather than extra measurable charges 
at infinity known as {\it dressing} \cite{Achucarro:1995nu}, 
has been known to be possible for many years, starting with the 
(unstable) coloured black holes \cite{Bartnik:1988am}, or the 
(stable) black monopoles, \cite{Lee:1991vy}, quantum hair,
\cite{Coleman:1991jf,Dowker:1991qe}, as well as the explicit vortex
hair \cite{Achucarro:1995nu}. 

In scalar tensor gravity, a number of  no-hair theorems 
(i.e., demonstrating that the scalar fields takes a constant value around 
an isolated black hole) have been proven, 
\cite{Bekenstein1996,Sotiriou2012,Faraoni2013} (although note the 
assumptions behind these theorems may not always correspond
to desired physical situations, \cite{Sotiriou:2013qea}).  The 
results of these theorems have been extended to binary black hole 
systems using perturbative, \cite{Yunes2012,Mirshekari2013}, and 
numerical, \cite{Healy2011}, calculations, demonstrating that 
dynamical spacetimes with two interacting black 
holes in scalar-tensor theory will be indistinguishable from general relativity. 
This might seem to imply that black hole systems will not be useful for 
constraining screened modified gravity, however, these theorems do not 
take into account cosmological backgrounds, in which the scalar field 
will typically be dynamical. A study of cosmologically evolving black holes 
with a canonical scalar field, \cite{Chadburn:2013mta}, shows that 
the scalar field does evolve on the event horizon of the black hole, 
though there is no evidence for an additional scalar charge. 
Indeed, there are several solutions in the literature which allow for
nontrivial scalar fields around black holes, 
\cite{Fonarev:1994xq,Clifton:2004st,Clifton:2006ug,Guariento:2012ri},
although to be fair, many are singular, or ``engineered''. 

This paper focuses on removing the requirement that the black hole 
exists in a vacuum.  This consideration is clearly relevant for astrophysical 
black holes that are typically observed in galaxies or galactic centres,
and can have energetic accretion discs.
In order to explore this question, we consider spherically symmetric 
distributions of matter around a black hole in screened modified gravity.
We perform a `probe' calculation, in that we explore the scalar profile
around the black hole without considering the corresponding modification
of the black hole geometry. We are therefore not looking at
issues of time dependent black holes, such as discussed in
\cite{McVittie:1933zz,Sultana:2005tp,Faraoni:2007es,Carrera:2008pi,Abdalla:2013ara},
as we will see nontrivial profiles even in the static case.

Strictly speaking, a non-uniform matter distribution will result
in a non-uniform scalar field profile due to the non-minimal 
coupling of the scalar to gravity. We expect that the complete 
picture will be a superposition of multipoles, with the dominant features being
encapsulated by the monopole, or spherically symmetric, behaviour. 
Indeed, if we were to discover that the spherically symmetric case
precluded the possibility of black hole hair, then it would be indicative
that black holes would not carry hair. Conversely, the discovery of 
non-trivial spherically symmetric scalar profiles would demonstrate
that black holes can indeed carry screened scalar hair, although 
the full solution would be more complicated and involved than the simple
picture presented here.

The layout of the paper is as follows: we introduce screened modified 
gravity in \textsection\ref{screenedMG}. We then study the effect of 
matter on the scalar
field profile in \textsection\ref{profile} under the assumption of spherical
symmetry and discuss observational implications for black hole properties in 
\textsection\ref{implications}. 

\section{Screened modified gravity}\label{screenedMG}

The basic idea of a screening mechanism is that either the mass of 
the additional scalar is dependent on the local energy density, or its
coupling to matter (or both). Thus, the field can be heavy or decoupled
in a dense environment such as our solar system
or galaxy, thereby giving no fifth force modifications to gravity, whereas
on large cosmological scales at very low densities, the field is much
lighter or can couple to matter and therefore give rise to modifications 
of the gravitational interaction.

We will explore three different models of screened modified gravity: 
the chameleon mechanism, \cite{Khoury2004},
which occurs when the mass of the scalar field, $m(\phi_0)$, 
is large enough to suppress the range of the scalar force;
the environmental dilaton, \cite{Brax2010}, where the coupling
function between the scalar and matter fields and the mass alter in 
dense regions; and the symmetron, \cite{Hinterbichler2011}, 
where the coupling function switches off in dense environments. 
These mechanisms can be modelled generically with the Einstein frame action
\be
\label{eframe_general}
S=\int d^4 x \sqrt{-g}\left[\frac{M_p^2}{2}R -\frac{1}{2}g^{\mu\nu}\partial_{\mu}\phi
\partial_{\nu}\phi -V(\phi)\right]+S_m\left[\Psi_i,A^2(\phi)g_{\mu\nu}\right].
\ee
Here $M_p^2 = 1/8\pi G$ is the Planck mass and $S_m$ is the action for 
matter fields (denoted generically as $\Psi_i$), which couple 
minimally to the Jordan frame metric, $\tilde{g}_{\mu\nu} = A^2(\phi)g_{\mu\nu}$.  
The details of a particular scalar-tensor theory are completely specified 
by the scalar potential $V(\phi)$ and the non-linear coupling function $A(\phi)$. 
We assume $A(\phi)$ is close to 1, writing
\be
\label{betadef}
A(\phi) = e^{\phi \beta(\phi)/M_p} \thickapprox 1 + \phi \frac{\beta(\phi)}{M_p}
\ee
As we are always in the low energy regime, we take $\phi \ll M_p$.
 
The modified `Einstein' equation is then written as
\be\label{EframeEeq}
G^{\mu\nu} = \frac{1}{M_p^2}\left(T_m^{\mu\nu} +T_{\phi}^{\mu\nu}\right)
\ee
where
\be
\begin{aligned}
T_m^{\mu\nu} &= 2\frac{\delta S_m}{\delta g_{\mu\nu}},\\
T_{\phi}^{\mu\nu} &= \nabla^{\mu}\phi\nabla^{\nu}\phi 
- g^{\mu\nu}\left(\frac{1}{2}g^{\alpha\beta}\nabla_{\alpha}
\phi\nabla_{\beta}\phi+V(\phi)\right)\,.
\end{aligned}
\ee

Using $T_m \equiv g_{\mu\nu}T_m^{\mu\nu}$, the scalar field equation is
\be
\label{scalareom}
\square \phi \equiv g^{\alpha\beta}\nabla_{\alpha}\nabla_{\beta}\phi 
= \frac{\partial V}{\partial \phi} - \frac{\partial \ln A}{\partial \phi}T_m.
\ee
Note that in the Einstein frame, the matter stress-energy tensor is not 
covariantly conserved,  rather, we have
\be
\label{EframedelT}
\nabla_{\mu}T_m^{\mu\nu} = -\nabla_{\mu}T_{\phi}^{\mu\nu} 
= T_m\frac{\partial \ln A}{\partial \phi}g^{\mu\nu}\partial_{\mu}\phi .
\ee
This can be interpreted as a `fifth force' on matter due to its interaction 
with the scalar field. For a non-relativistic particle, this takes the form 
\be
\label{Egeo_nonrel}
\ddot{\bf x} = -\frac{\partial \ln A}{\partial\phi} \mathbf{\nabla}\phi\,.
\ee

It is convenient to define $\rho \equiv -A^{-1}T_m$, which is a 
conserved density for non-relativistic distributions 
of matter, \cite{Brax2012a}, in the Einstein frame.  Using this 
formalism, the scalar equation of motion is
\begin{equation}\label{fieldeq}
\square\phi = \frac{\partial}{\partial \phi}\left[V(\phi) 
+  (A(\phi)-1)\rho\right]\equiv
\frac{\partial  V_{\text{eff}}(\phi,\rho)}{\partial\phi}.  
\end{equation} 
The fact that dynamics of $\phi$ are governed by a density-dependent effective 
potential $V_{\text{eff}}(\phi,\rho)$ is the source of the screening behaviour for 
chameleons, environmentally dependent dilatons, and symmetrons.  

\subsection{Chameleons}\label{ch_model}

\begin{figure}
\centering
\includegraphics[scale=0.6]{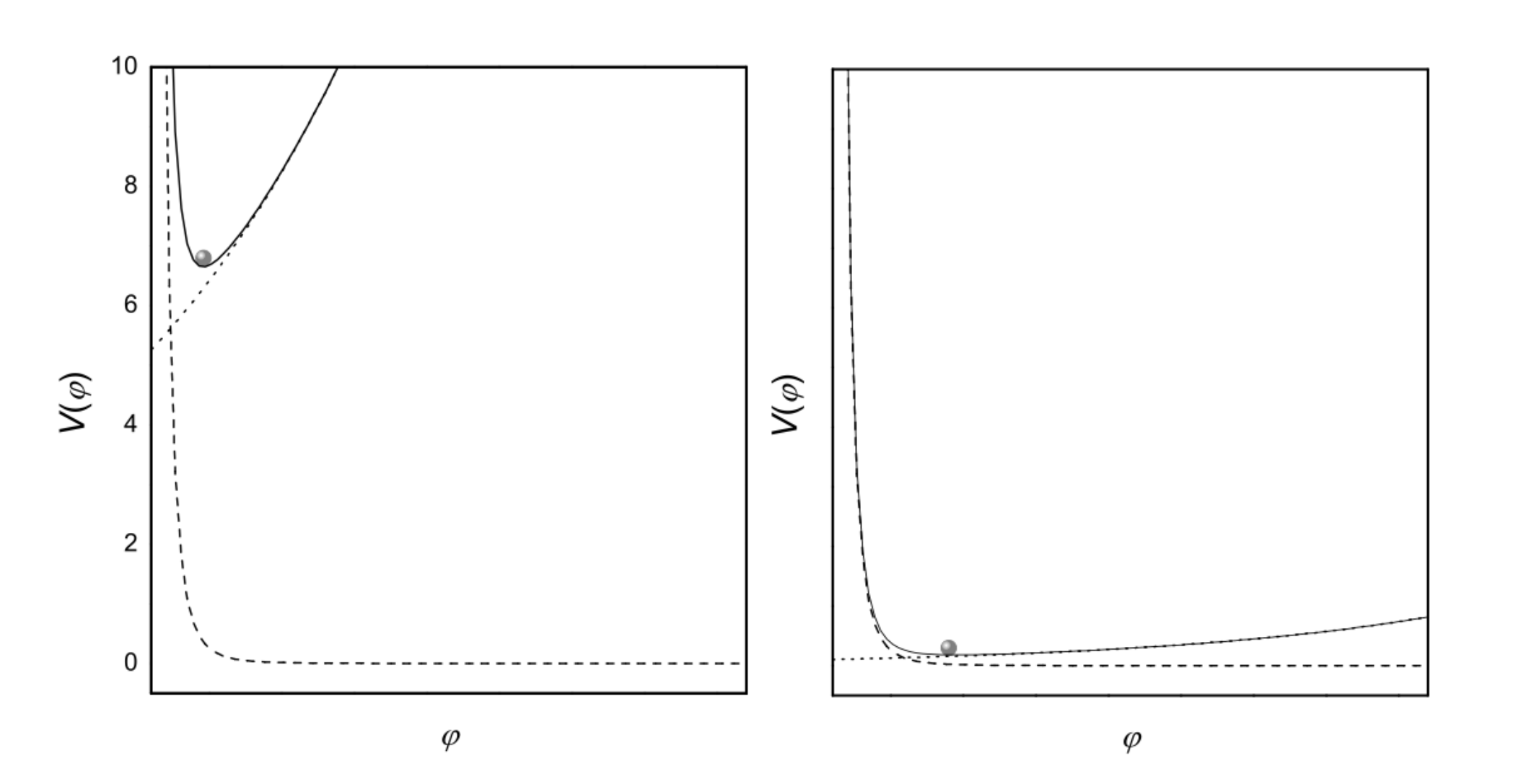}
\caption{Plots illustrating the chameleon screening mechanism,
\cite{Brax:2013mua}. The dashed, dotted and solid lines are the 
bare potential $V(\phi)$, the coupling function, and $V_{\text{eff}}$
respectively. \textit{Left Panel:} in high density regions, the 
minima of $V_{\text{eff}}$ is close to $\phi=0$ and $\vec{\nabla}\phi$ 
and hence the fifth force is small. \textit{Right Panel:} in low density 
regions the fifth force can be non-trivial.}
\label{fig:champlots1}
\end{figure}

Chameleon models contain a scalar field whose mass is an increasing 
function of density, which causes the range of any fifth force to be suppressed 
in dense regions~\cite{Khoury2004}. This means chameleon scalars 
are screened in dense environments like the solar system, and can evade 
local observational constraints while still having significant effects
on cosmological scales. In these models, it is typically assumed 
that $\beta(\phi)$ is nearly independent of $\phi$ throughout the 
relevant field range, and so can be treated as a constant parameter. 
The coupling function is therefore taken to be
\be 
A(\phi) = e^{\beta\phi/M_p}\,.
\ee
A typical chameleon potential is
\be\label{chV1}
V(\phi) = M^{4+n} \phi^{-n}=  V_0 \phi^{-n},
\ee
where $n\geq 1$ is an integer of order one, and we define 
$V_0\equiv M^{4+n}$ to simplify notation. Keeping only the leading 
order term from the coupling function, we 
see that the effective potential is
\be
\Veff(\phi,\rho)\approx \frac{V_0}{\phi^{n}} + \frac{\rho\beta\phi}{M_p},
\ee
which is minimised at
\be
\phi_{\text{min}}^{n+1} =\frac{nV_0M_p}{\rho\beta}.
\ee
The mass of small fluctuations of the field around this minimum is
\be\beal
m^2(\rho) &=\left.\Veff(\phi,\rho),_{\phi\phi}\right|_{\phi_{\text{min}}} \\
&\approx \frac{\rho\beta}{M_p} \left[
(n+1)\left(\frac{\rho\beta}{nV_0M_p}\right)^{\frac{1}{n+1}} 
+\frac{\beta}{M_p}\right]\label{ch_masseq}
\eeal\ee
which, as required, increases monotonically with $\rho$.

Current constraints on chameleon models come from laboratory, 
cosmological, and astrophysical tests.  Fifth force constraints from 
E\"{o}t-Wash torsion-balance experiments~\cite{Will2005} give the 
bound $M\lesssim 10^{-3}\,\text{eV}$, assuming $\beta$ and $n$ 
are of order one~\cite{Jain2010,Upadhye2012,Brax2012a,Mota2007}.
Demanding that the Milky Way be  screened\footnote{i.e.\ the effect 
of the scalar fifth force on a test particle be negligible compared 
to the gravitational force, which from \eqref{Egeo_nonrel} implies 
$d\ln A(\phi)/d r \le d\Phi_N/dr$, where $\Phi_N$ is the Newtonian 
potential}, gives a lower bound 
for the mass of the chameleon at cosmological densities, 
$m_{\text{cosm}}\gtrsim 10^3 H_0$ or equivalently, $
m_{\text{cosm}}^{-1} \lesssim 1\,\text{Mpc}$,
\cite{Khoury2012, Brax2004a}. Laboratory constraints restrict the 
mass at terrestrial densities to be  $m_{\oplus}^{-1} \lesssim 
50\,\mu\text{m}$~\cite{Kapner:2006si}.

We can use \eqref{ch_masseq} to translate constraints on the 
chameleon's cosmological or terrestrial mass to an estimate in 
regions with arbitrary density.  
Assuming $\beta\sim\mathcal{O}(1)$, we note that if we set $V_0$ 
by the dark energy scale, $M\sim 10^{-3}\,\text{eV}$, as indicated by
the limits, then the first term 
inside of the square brackets in \eqref{ch_masseq} will always 
dominate for $\rho\gtrsim \rho_{\text{cosm}}\sim H_0^2 M_p^2$.  
This allows us to approximate the ratio of the chameleon's 
effective mass at two different densities as
\be
\label{ch_massrel}
\frac{m_a}{m_b} \sim \left(\frac{\rho_a}{\rho_b}\right)^{\frac{1}{2}
\left(\frac{n+2}{n+1}\right)}.
\ee

We can therefore see that the strongest constraint comes
from the laboratory experiments, and this is the bound we will use
for the chameleon Compton wavelength. 
In our calculations it will be important to know whether the Compton 
wavelength of the chameleon is large or small compared to the size of the 
black hole.  To facilitate this comparison, Table~\ref{ch_masstable} gives a 
summary of the bounds for $m^{-1}(\rho)$ at the 
various densities we are interested in, obtained by rescaling the
laboratory upper bound.

\begin{table}\centering
\caption{Order of magnitude bounds for the chameleon 
wavelength}
\vskip 3mm
\label{ch_masstable}
\begin{tabular}{ || l | c | r  c c ||}
\hline\hline
Environment & Density & \multicolumn{2}{c}{Compton wavelength
upper bound}& \\
& & ~~~~$n= 1$ & $n$ large&\\
\hline 
Earth &  $\rho_{\oplus}\sim10^{29}\rho_{\text{cosm}}$
&$10^{-5}\,\text{m}$ &$10^{-5}\,\text{m}$&\\
Accretion disc & $10^{-8}\rho_{\oplus}$ &$10\,\text{m}$ 
&$0.1\,\text{m}$ &\\
Galaxy & $10^6\rho_{\text{cosm}}$ 
& $10^{12}\,\text{m}$&$10^{7}\,\text{m}$& \\
\hline  \hline
\end{tabular}\\
\end{table}

A solar mass black hole has a Schwarzschild radius of $\sim 10^3 \,\text{m}$, 
while a supermassive black hole has $R_s\sim 10^{9-12}\,\text{m}$.  
Referring to Table \ref{ch_masstable}, we see that the Compton wavelength 
of the chameleon will always be very small compared to the characteristic 
length scale of a black hole surrounded by an accretion disk. However, if 
there is no accretion disk and the black hole is simply surrounded by the 
ambient galactic density, the chameleon wavelength while potentially 
short compared to the size of a supermassive black hole, will be long 
compared to that of a stellar mass black hole. Interestingly, for the
particular chameleon model we explore most fully in \textsection \ref{ch_profile},
which has $n=4$, the Compton wavelength limit at galactic densities is
of the same order ($10^9$m) as the radius of a typical supermassive
black hole.

\subsection{Environmentally dependent dilatons}\label{dil_model}

Environmentally dependent dilaton models rely on a generalisation of the
Damour-Polyakov mechanism, which arises in the strong coupling limit
of string theory~\cite{Damour:1994zq}. 
Their screening behaviour is a consequence of the fact 
that their coupling function has a minimum at some $\phi_d$,
\be
A(\phi) = 1+\frac{a_2}{2M_p^2}(\phi - \phi_d)^2 +\dots
\ee
or
\be
\beta(\phi) \equiv M_p \frac{\partial \ln A(\phi)}{\partial \phi} 
\thickapprox \frac{a_2}{M_p} (\phi-\phi_d)\,.
\label{dil_betadef}
\ee
The dilaton self-interaction potential is 
\be
V(\phi) =A^4(\phi)M^4 e^{-\frac{\phi}{M_p}} \equiv A^4(\phi) 
V_0 e^{-(\phi-\phi_d)/M_p}
\ee
where again we define $V_0\equiv M^4 e^{-\phi_d/M_p}$ to simplify notation.
\begin{figure}
\centering
\includegraphics[scale=0.55]{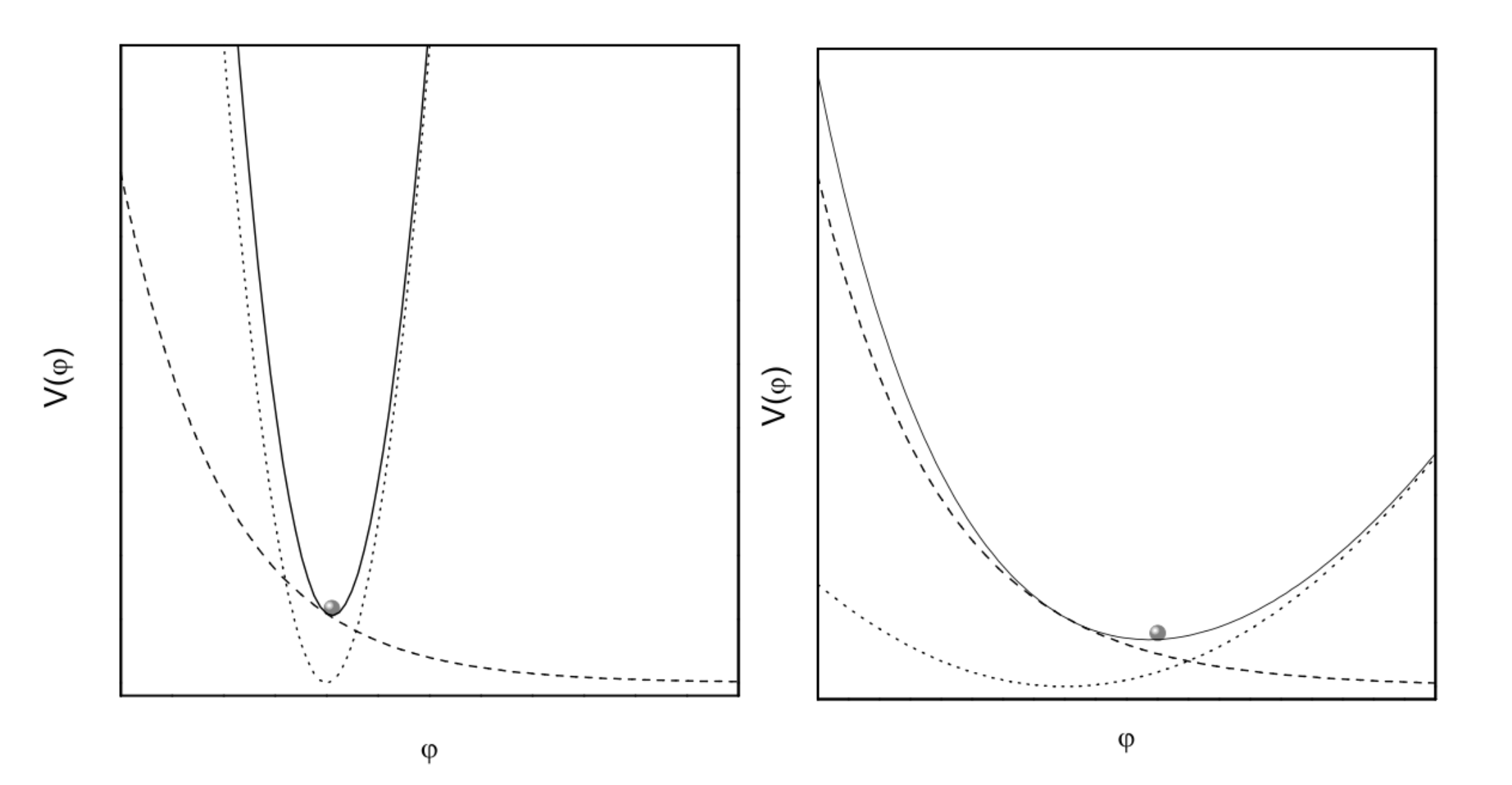}
\caption{Plots illustrating the dilaton screening mechanism~\cite{Brax:2012nk}. 
The dashed, dotted and solid lines are respectively the bare potential $V(\phi)$, 
the coupling function and $V_{\text{eff}}$. \textit{Left Panel:} in high density 
regions, the minima of $V_{\text{eff}}$ is where the coupling strength vanishes 
and hence the fifth force is small. \textit{Right Panel:} in low density regions 
the fifth force can be non-trivial.}
\label{fig:dilplots1}
\end{figure}

There is a caveat in that the action for dilatons contains a non-canonical 
kinetic term: $k^2(\phi)(\nabla\phi)^2$. The function $k(\phi)$ depends 
on the parameters of the particular string theory, but assuming that 
$\phi/M_p\ll 1$ it will generally
take the form
\be\label{dil_kdef}
k(\phi)\approx \lambda^{-1}\sqrt{1+3\lambda^2\beta^2(\phi)}
\ee
where $\lambda$ typically falls in the range $\mathcal{O}(1)\lesssim 
\lambda \lesssim \frac{M_p}{M_s}$. Here, $M_s$ is the string energy scale
which we will take to be of order $10^{-1} - 10^{-2} M_p$. Thus our
equation of motion is
\be
\begin{aligned}
\Box \varphi  &= k \Box \phi + k' \left ( \nabla\phi \right)^2
= \Veff,_{\varphi}(\varphi,\rho) = k^{-1}(\phi) \Veff,_{\phi}(\phi,\rho)\\
&\simeq k^{-1}(\phi) \frac{V_0}{M_p} \left [ (4\beta(\phi) -1) 
A^4(\phi) e^{-(\phi-\phi_d)/M_p} + \beta(\phi) A(\phi) 
\frac{\rho}{V_0} \right]
\end{aligned}
\ee
where the redefined scalar, $\varphi$, is defined via $d\varphi = k(\phi)d\phi$.

The effective potential will be minimised when 
\be\label{dil_betaeq}
\beta_{\text{min}}\equiv\beta(\phi_{\text{min}})
= \frac{V_0}{4V_0 + \rho A^{-3} e^{(\phi_{\text{min}}-\phi_d)/M_p}} 
\simeq \frac{V_0}{\rho+4V_0},
\ee
for $\phi_{\text{min}}\equiv \phi(\varphi_{\text{min}})$.  Thus, for large 
matter density, $\beta$ is suppressed and the scalar field will decouple 
from matter. We also note that  $0<\beta\leq\frac{1}{4}$, so $\beta$ is 
at most order unity for dilatons. 

The mass of small fluctuations about $\varphi_{\text{min}}$ is
\be
\begin{aligned}
m_{\varphi_{\text{min}}}^2 
&=\left.\Veff,_{\varphi\varphi}\right|_{\varphi_{\text{min}}} = 
\left.\frac{\Veff,_{\phi\phi} }{k^2}\right|_{\phi_{\text{min}}} \\
&\approx \frac{\lambda^2 V_0}{M_p^2}
\frac{(12\beta^3_{\text{min}}-6\beta^2_{\text{min}}+a_2)}
{\beta_{\text{min}}(1+3\lambda^2\beta^2_{\text{min}})}.
\end{aligned}
\ee
As with chameleons, the requirement that the Milky Way be screened 
places a lower bound on the mass of the dilaton at cosmological densities 
of $m_{\text{cosm}}\geq 10^3 H_0$~\cite{Brax2012}.  
This translates to the requirement that $\lambda^2 a_2 \gtrsim 10^5$. 
We  note that because  $0<\beta\leq \frac{1}{4}$ and $a_2\gg 1$, the 
dilaton's mass is dominated by the final term in the numerator, 
$m_{\varphi_{\text{min}}}^2 \sim \lambda^2 a_2 (\rho+4V_0)/M_p^2$.  
Saturating the bound on $\lambda^2a_2$ gives an upper bound 
of ${\cal O}(10^{22})$m for the Compton wavelength
of the dilaton in vacuum, which is always much larger than any black hole
of astrophysical or cosmological interest. Meanwhile, in a denser 
region surrounding a black hole, the coupling function is markedly
damped, see \eqref{dil_betaeq}, resulting in the dilaton effectively 
decoupling from matter and being fixed very close to $\phi_d$.

\subsection{Symmetrons}\label{sym_model}

Like dilatons, symmetrons exhibit screening behaviour because their 
coupling to matter goes to zero in dense regions,
\cite{Hinterbichler2011,Upadhye2012b,Clampitt2011,Khoury2010}.  
They have a symmetry breaking potential and a quadratic coupling function,
\be
V(\phi) = -\frac{\mu^2}{2}\phi^2 + \frac{\lambda}{4}\phi^4,\quad\quad\quad 
A(\phi) = 1+\frac{a_2}{2M_p^2}\phi^2.
\ee
This gives the effective potential symmetry breaking properties which 
depend on the local matter density,
\be
\Veff(\phi,\rho) = \left(\frac{\rho a_2}{M_p^2} -
\mu^2\right)\frac{\phi^2}{2} +\lambda\frac{\phi^4}{4}.
\ee
In regions with $\rho< \mu^2 M_p^2/a_2$, the symmetron will acquire 
a non-zero vacuum expectation value (VEV) of  $\phi_{\text{min}}
=\pm \frac{\mu}{\sqrt{\lambda}}$. In these regions, the field couples 
to matter through $\beta(\phi)\approx a_2\phi_{\text{min}}/M_p$ and 
will produce a fifth force.  When  $\rho$ is large enough to restore symmetry, 
the VEV of $\phi$ vanishes, hence $\beta(\phi)\thickapprox 0$ thus
decoupling the symmetron from matter. 
\begin{figure}
\centering
\includegraphics[scale=0.55]{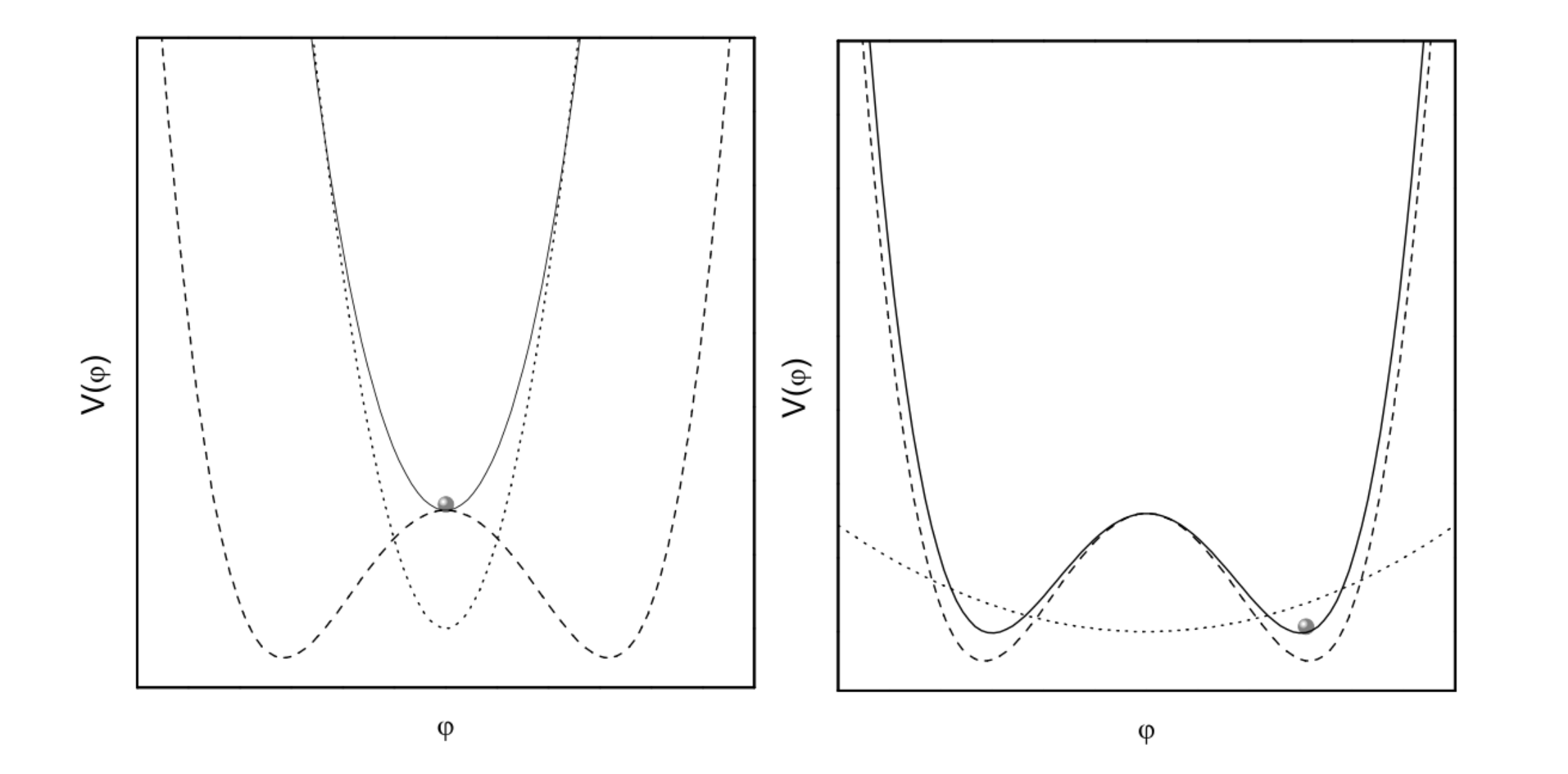}
\caption{Plots illustrating the symmetron screening mechanism,
\cite{Brax:2012nk}. The dashed, dotted and solid lines are the bare 
potential $V(\phi)$, the coupling function, and $V_{\text{eff}}$ respectively. 
\textit{Left Panel:} in high density regions, the minima of $V_{\text{eff}}$ 
is where the coupling strength vanishes and hence the fifth force is small. 
\textit{Right Panel:} in low density regions the fifth force can be non-trivial.}
\label{fig:symplots1}
\end{figure}

Note that though the screening behaviour does not depend on it, the 
effective mass of the symmetron will be different in regions of different density:
\begin{align}\label{sym_masseq}
\text{for}\;\rho> \mu^2 M_p^2/a_2 \quad \quad m^2
&\equiv\left.\Veff,_{\phi\phi}(\phi,\rho)
\right|_{\phi=0}  = \frac{\rho a_2}{M_p^2}-\mu^2\\
\text{for}\;\rho< \mu^2 M_p^2/a_2\quad \quad m^2
&\equiv\left.\Veff,_{\phi\phi}(\phi,\rho)
\right|_{\phi=\pm\frac{\mu}{\sqrt{\lambda}}}  = 2\mu^2
\end{align}
In order for symmetrons to produce long-range modifications of gravity, 
the effective potential should have spontaneously broken symmetry 
at cosmological densities, i.e., $\rho_{\text{cos}}\sim H_0^2 M_p^2
\lesssim \mu^2 M_p^2/a_2$ and requiring the Milky Way to be screened imposes 
$a_2\gtrsim 10^{8}$, \cite{Hinterbichler2011}, hence
$\mu\gtrsim 10^4 H_0$ ($\mu^{-1}\lesssim 0.4\,\text{Mpc}$).  
Additionally, for the symmetron to have effects comparable to gravity in 
low-density regions, we need $\beta(\phi_{\text{min}})\sim\mathcal{O}(1)$,
which then requires $\lambda$ to be very small: $\lambda
\gtrsim 10^{24}(H_0/M_p)^2\sim 10^{-96}$.

We can estimate an upper bound on the wavelength of the symmetron 
at arbitrary density by combining these constraints with \eqref{sym_masseq}.  
For regions with density greater than about $10^{-6}\,\text{eV}^4$, 
the $\mu^2$ term can be neglected, giving
\be
m^{-1}(\rho)\lesssim \sqrt{\frac{\rho_{\text{cos}}}{\rho}}
\times {\cal O}(10^{22})\,\text{m}.
\ee
Thus at physically realistic environmental densities, 
it is reasonable to assume that for symmetrons $m^{-1}$ will be 
large compared to black hole systems.

\section{Scalar profile of a black hole surrounded by matter}\label{profile}

Our goal is to study the scalar field profile induced by a non-uniform matter distribution around a black hole and to estimate the 
magnitude of any scalar gradients which could potentially affect observable 
properties. Our approach will be to look for a static solution to the
scalar field equation \eqref{fieldeq} on the Schwarzschild black hole 
geometry, together with a non-uniform 
matter distribution which is assumed (as with the scalar) not to back-react
upon the geometry to this order. There are two aspects to this assumption.
Firstly, on spatial scales over which the black hole curvature is significant,
it requires that the curvature induced by the matter/scalar field be
subdominant to that of the black hole. This translates roughly to the requirement
\be\label{Tphireq}
|T_{\phi}|,\, \rho \ll \frac{M_{BH}}{R_S^3} \approx  \frac{M_p^6}{M_{BH}^2}.
\ee
Secondly, although on cosmological timescales we do expect the 
matter and scalar to have an effect on the spacetime geometry, 
this will involve a time evolution of order the Hubble scale, thus for 
the purposes of exploring observational consequences this is very 
much subdominant to local environmental timescales for the black hole, 
or reaction of the scalar field to the density profile\footnote{In 
\cite{Chadburn:2013mta}, the time dependence of the back-reacted 
cosmological scalar plus black hole was shown to be at a timescale
determined by the cosmological expansion, not the black hole light
crossing time.}. Thus, the assumption of staticity is reasonable for timescales
$\tau \ll H_0^{-1}$, which is safely the case for our astrophysical black holes.

One objection that might be raised to this approach is that the ``no hair''
theorems preclude any nontrivial scalar profile. The relevant no-hair
theorem was explored by Sotiriou and Faraoni, \cite{Sotiriou2012},
for the case of static and vacuum solutions. The essence of their 
argument is to take the scalar equation of motion (shown here for the
Schwarzschild background), to multiply by $V_{,\phi}(\phi) \sqrt{g}$ 
and integrate:
\be
\int_{2GM}^\infty \left \{ r^2 V_{,\phi}^2 
+ r(r-2GM) \phi' \frac{d}{dr} \left ( V_{,\phi} \right ) \right \}
= \left [V_{,\phi} r(r-2GM) \phi' \right ]^\infty_{2GM} = 0.
\label{nohairint}
\ee
Clearly, if $V_{,\phi\phi} \geq 0$, as is the case with a wide range of
physically relevant potentials, then the only possibility is that the integrand
on the LHS is identically zero, i.e., $\phi' \equiv 0$, $\phi = \phi_{\rm min}$.
Although at first sight the potentials we are considering appear to 
satisfy this constraint, we must be careful, as it is the effective potential
that is the relevant quantity, and we are looking at a non-uniform environment
where the matter density, $\rho$, jumps from being roughly zero to the ambient
galactic or local accretion disc value. Thus, although $r^2 V_{,\phi}^2$
is positive definite, the derivative of $V_{,\phi}$ with respect to $r$ 
contains a delta function, coming from the derivative of $\rho$. 
Combining this with the intuition that $\phi$, if nontrivial, will tend to 
roll towards large values near the black hole horizon, we see that
the second term in the integrand can potentially be very large and
negative, thus ruling out a simple ``no-hair'' proof, and opening the 
possibility of a nontrivial scalar profile.

Our setup is motivated by a 
physical picture of an astrophysical black hole, typically 
located within some larger distribution of matter. 
Although astrophysical black holes will be rotating, for the purpose
of establishing whether or not a nontrivial scalar profile is possible,
it will suffice to consider a purely monopole spherically symmetric
set-up, in which the black hole is descibed by
the Schwarzschild metric:
\be
ds^2 = -\left(1-\tfrac{R_s}{r}\right)dt^2 + \left(1-\tfrac{R_s}{r}\right)^{-1}dr^2 
+r^2 d\Omega^2 \,,
\ee
(denoting $R_s = 2GM$ for clarity), and the density profile by
\be\label{rhodef}
\rho(r)\equiv \begin{cases}
0 & R_s<r<R_0 \text{ (Region I)}\\
\rhostar  & r>R_0\text{ (Region II)}
\end{cases}
\ee
The motivation for this profile is that the larger distribution of matter in
which the black hole sits will be characterised by a density $\rhostar$ 
(taken to be constant in \eqref{rhodef}), which is assumed to vary slowly on 
length scales comparable to the size of the black hole.  
Very close to the black hole however, we expect an approximately 
empty inner region, motivated by the fact that all black holes have 
an innermost stable circular orbit (ISCO -- e.g.\ at $3R_s$ for the 
Schwarzschild black hole), inside of which
all massive particles fall into the black hole on a relatively short time-scale.
We therefore treat the density inside some inner radius, $R_0$ as
being roughly zero. Our matter profile \eqref{rhodef} can thus be viewed  
as a crude model of either a {\it{spherically symmetric}} accretion ``disk'' 
or  a galactic ``halo'' where the 
matter inside $R_0$ has fallen into the black hole.

Using this model, and taking $\phi = \phi(r)$, the scalar field equation 
becomes an ordinary differential equation:
\be\label{generaleom}
\square\phi =  \frac{1}{r^2} \frac{d}{dr}\left[
r^2\left(1-\frac{R_S}{r}\right)\frac{d\phi}{dr}\right]
=\frac{\partial  V_{\text{eff}}(\phi,\rho)}{\partial\phi}. 
\ee
The  external matter distribution sets the asymptotic boundary 
condition $\phi \to \phistar$, where 
$\left.\Veff(\phi,\rhostar),_{\phi}\right|_{\phistar} =0$. 
Note that we typically expect $\phi(r)$ to approach this minimum over a 
length scale characterised by $\mstar^{-1}$, where $\mstar\equiv
\left.\Veff(\phi,\rhostar),_{\phi\phi}\right|_{\phistar}$. 

We now solve this scalar equation of motion for chameleons 
\textsection\ref{ch_profile}, dilatons \textsection\ref{dil_profile}, and 
symmetrons \textsection\ref{sym_profile}, respectively.
Note that although we expect $R_0 \simeq R_{\rm ISCO} = 3R_s$
for the Schwarzschild black hole, to make our analytic calculations
tractable we will often use the approximation $R_s/R_0\ll 1$. 
While the actual value of $1/3$ means there will be quantitative 
inaccuracy to the analytic expressions, we nonetheless expect that 
the qualitative picture emerging from our analytic results will be correct. 

\subsection{Chameleon profile}\label{ch_profile}
 
We first present an analytic approximation to the Chameleon 
scalar profile, and derive the horizon value of the field in two limits: 
where the scalar Compton wavelength is either very large or very small 
compared to the size of the black hole.  
As we saw in Section~\ref{ch_model}, if we take $\rhostar$ to be the 
density of an accretion disk, $\mstar R_0$ will always be large ($\gg 1$). 
If $\rhostar$ is the density of a galactic halo the chameleon will have 
$\mstar R_0 \ll 1$ for stellar mass black holes, 
and $\mstar R_0\lesssim 1$ for supermassive black holes.  Thus, both limits will 
potentially be relevant for chameleons.

It proves convenient to rewrite the chameleon equation of motion in terms of 
dimensionless variables:
\be
\hat{\phi} = \frac{\phi}{\phi_\star}\;, \;\;
x=\frac{r}{R_s}\;, \;\;
{\hat m}^2 = m_\ast^2 R_s^2 
= (n+1) \frac{\rho_\ast \beta R_s^2}{M_p \phi_\ast}
\ee
giving
\be
\hat{\phi}'' + \frac{2x-1}{x(x-1)} \hat{\phi}' = \frac{x}{(x-1)} 
\frac{{\hat m}^2}{(n+1)}
 \left [ \Theta[x-x_0]
- \frac{1}{\hat{\phi}^{n+1}}\right]
\label{chameq}
\ee

The physical set-up is that we have a dense extended region (II) in
which the chameleon will be held essentially constant at $\phi_\ast$.
Nearer to the black hole, we have a region of vacuum (region I) in
which the chameleon is allowed to roll freely and is only restricted by
the dimension of region I. For a low mass chameleon (large
Compton wavelength with respect to the black hole) we do not expect
the chameleon to change much from its asymptotic value, hence we
can perform an analytic approximation assuming a small change
in $\hat\phi$. For large mass chameleons however, we do expect
a rather sharp and rapid response to the vacuum region, and for the
chameleon to have something analogous to a thin shell with
a power law behaviour, commensurate with the rolling of the vacuum
potential. Thus, for our analytic approximation we also use a
different expression in region I for large masses.
In both cases in region II however, the field will fall off to its asymptotic 
value, and we expect ${\hat\phi} \simeq 1 + \delta{\hat\phi}$, where
\be
\delta{\hat\phi} \simeq C \frac{e^{-{\hat m} (x-x_0)}}{x^{1+{\hat m}/2}}
\label{chregIIapp}
\ee
for some constant $C$. 
\medskip

\noindent $\bullet$ ${\hat m}x_0\ll1$

For a long range chameleon field, we expect that $\hat\phi$ will not
vary much, and assume the change in $\hat\phi$ in region I is 
dominated by the geometry, i.e.\ 
\be
\left [ x(x-1) \hat{\phi}'\right]' = -
\frac{{\hat m}^2 x^2}{(n+1)\hat{\phi}_h^{n+1}} 
\left ( 1 + {\cal O} (\delta {\hat\phi}/{\hat \phi}_h )\right)\,,
\ee
which gives the solution:
\be
{\hat \phi} = {\hat\phi}_h - \frac{{\hat m}^2}{6(n+1) {\hat\phi}_h^{n+1}}
[x^2 + 2x + 2\ln x-3]\,.
\label{chregIapp}
\ee
Matching the solutions at $x_0$ with the asymptotic form \eqref{chregIIapp}
gives
\be
\begin{aligned}
{\hat \phi}_h &= 1 + \frac{{\hat m}^2}{6(n+1){\hat\phi}_h^{n+1}}
\left [ x_0^2 + 2 x_0 + 2 \ln x_0 -3
+ \frac{4x_0^2 + 4 x_0 +4}{2 {\hat m}x_0 + 2 + {\hat m}}\right ]\\
C &= \frac{{\hat m}^2}{3(n+1) {\hat\phi}_h^{n+1}}
\frac{x_0^{1+{\hat m}/2}(x_0^2+x_0+1)}{{\hat m}x_0+1+{\hat m}/2}\;.
\end{aligned}\label{chamdata}
\ee

Writing ${\hat\phi} = 1 + \delta{\hat\phi}$, and expanding to leading order
gives:
\be
{\hat\phi} \simeq 1+ \begin{cases}
\frac{{\hat m}^2}{6(n+1)}
\left [ 3x_0^2 -x^2 + 4 x_0 -2x + 2 +2 \ln \frac{x_0}{x} \right ]& x<x_0\\
\frac{{\hat m}^2}{3(n+1)}\left ( x_0^2 + x_0 + 1 \right)
\frac{x_0}{x} e^{-{\hat m} (x-x_0)}& x>x_0
\end{cases}
\label{analappchamsm}
\ee

We are mainly interested in the horizon value of the chameleon field,
and how this differs from the asymptotic value, as this indicates the
impact of the black hole on the local scalar profile, and we can read
this off to leading order in ${\hat m}x_0$ as: 
\be
{\hat\phi}_h \approx 1+ \frac{{\hat m}^2 x_0^2}{2(n+1)}\,,
\ee
or,
\be\label{chl_dphih}
\delta\phi_h \approx \frac{\rhostar\beta R_0^2}{2M_p}.
\ee
Here we see that $\delta\phi_h$ increases with the coupling 
function $\beta$, the density of the local environment and the
range in which the chameleon can roll ($R_0$).

\medskip

\noindent $\bullet$ ${\hat m}x_0\gg 1$

In the case that the chameleon is short range, we can use the same
simple approximation presented above, although without expanding
\eqref{chamdata} near ${\hat \phi}=1$. This approximation should
give a good order of magnitude estimate for ${\hat \phi}_h$, however
for later purposes we want a better approximation to the field
profile so we can estimate $\phi'$. Since we expect the key 
feature of the profile to be the rapid roll of the chameleon near the
boundary of the two regions, we focus on the change
in $\hat\phi$ being dominated by the potential:
\be
{\hat\phi}'' \simeq - \frac{{\hat m}^2}{(n+1)} \frac{1}{{\hat\phi}^{n+1}}
\ee
which is solved by
\be
{\hat\phi} \simeq 1+ \left [ \frac{{\hat m}^2 (n+2)^2 (x_1-x)^2} 
{2n(n+1)}\right]^{1/(n+2)}\,.
\ee
Meanwhile, matching to \eqref{chregIIapp} at $x_0$ gives:
\be
{\hat\phi} \simeq 1+ \begin{cases}
\left [ \frac{{\hat m}^2 (n+2)^2 }{2n(n+1)}\right]^{\frac{1}{n+2}} 
\left (x_0-x + \frac{2}{(n+2){\hat m}}\right)^{\frac{2}{n+2}}& x<x_0\\
\left [ \frac{ 2}{n(n+1)}\right]^{\frac{1}{n+2}} \left (\frac{x_0}{x}\right)^{1+{\hat m}/2}
e^{-{\hat m} (x-x_0)}& x>x_0
\end{cases}
\label{analappchamlg}
\ee

This profile captures a rapid transition to large near horizon values,
although it does not solve the equations of motion at the horizon, as
it has been tailored to the variation near $x_0$. In spite of this,
as we will see from the numerical work, it does indeed capture the essential
features of the field profile (see next subsection). Indeed, both this
expression, and the simpler geometry dominated approximation data
\eqref{chamdata} give the same dependence of the horizon value on 
$x_0$ and ${\hat m}$ to leading order in $x_0$
\be
{\hat\phi}_h \approx {\hat\phi}_c \left ( \frac{{\hat m}^2 x_0^2 }{(n+1)}
\right)^{\frac{1}{n+2}}\,,
\ee
where $ {\hat\phi}_c^{n+2} = (n+2)^2/(2n)$ for the potential dominated
expression, and $1/6$ for the geometry dominated expression. 
Re-expressing in terms of the dimensionful variables gives
\be\label{ch_phih}
\phi_h \propto \left((n+2)^2 V_0 R_0^2\right)^{\frac{1}{n+2}}.
\ee

This result shows how the value of the chameleon is dependent on the
parameters of the model: Making the chameleon self-interaction potential 
$V(\phi)$ steeper by lowering $n$ or raising $V_0$ will drive $\phi_h$ to 
larger values, as will increasing $R_0$.

\medskip

In both cases, the field is close to its asymptotic value in region II,
and rolls to larger values of $\phi$ in region I. For the long range
chameleon, we expect our analytic approximation to be very good
and accurate up to sub-leading dependence on $x$ of order 
${\hat m}/x_0$. For the short range chameleon, we could not find
a simple analytic expression that worked throughout region I: the 
potential dominated expression should work well near $x_0$, however,
the effect of the nearing event horizon should modify this profile
once we are at smaller $x$. Both approximations gave the same
form for the horizon value of the chameleon however, therefore
we expect the actual profile to have features of both, and certainly
to be nontrivial!

For both small and large mass chameleons,
increasing the size of the empty region around the black hole 
(thus giving the chameleon more `space' to roll) increases $\phi_h$.  
We can read this as a rough restatement of a no-hair theorem, noting 
that as we take $R_0\rightarrow\infty$ the 
requirement that the scalar stay finite is violated. 

\begin{figure}
\centering
\includegraphics[scale=0.5]{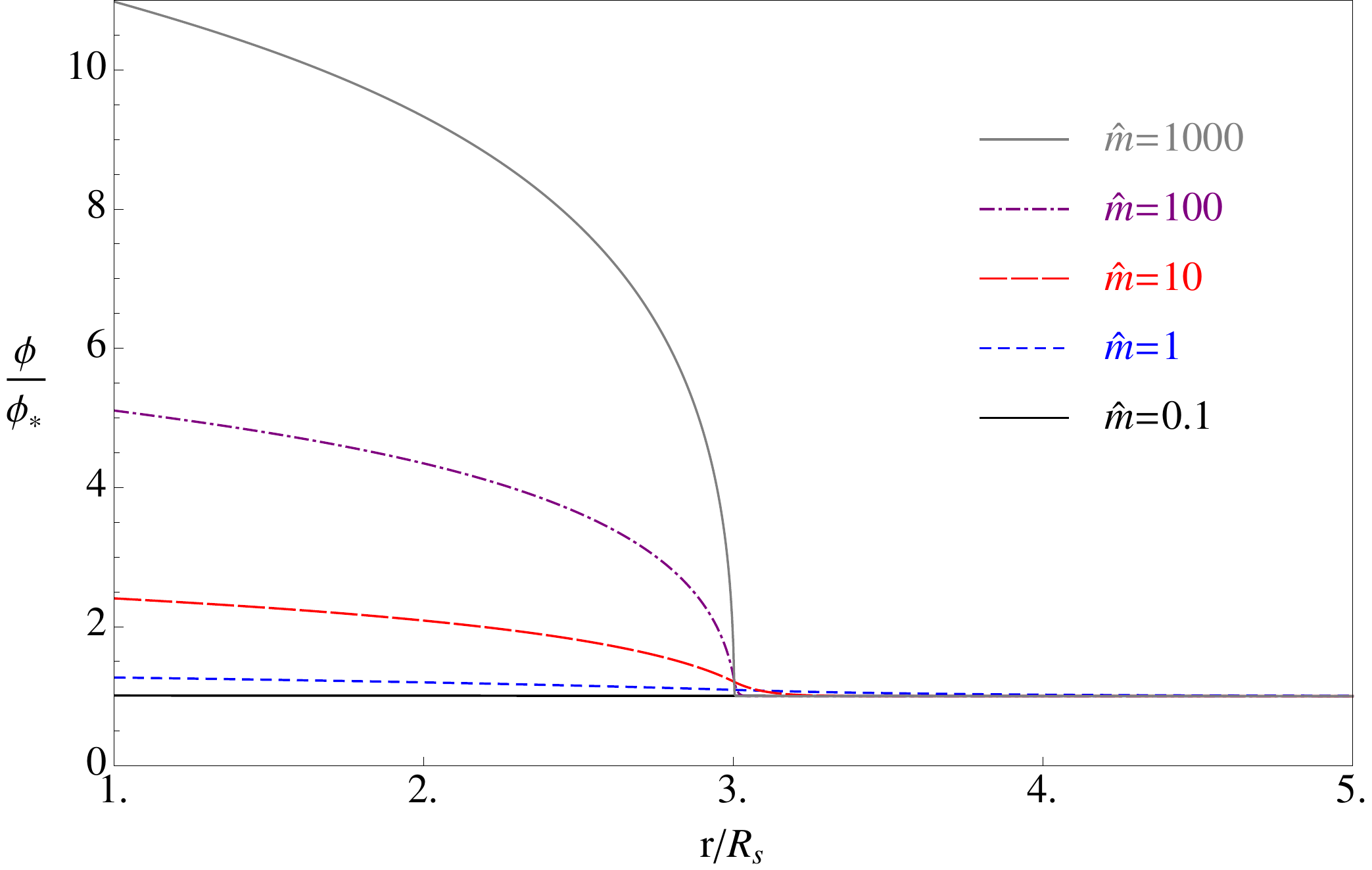}
\caption{Plots of numerical solutions for the chameleon field for a 
range of values of the chameleon mass ${\hat m} = m_\star R_s$. 
The differing behaviour for the large and low mass solutions is 
demonstrated.}
\label{fig:champlots}
\end{figure}

\begin{figure}
\centering
\includegraphics[scale=0.32]{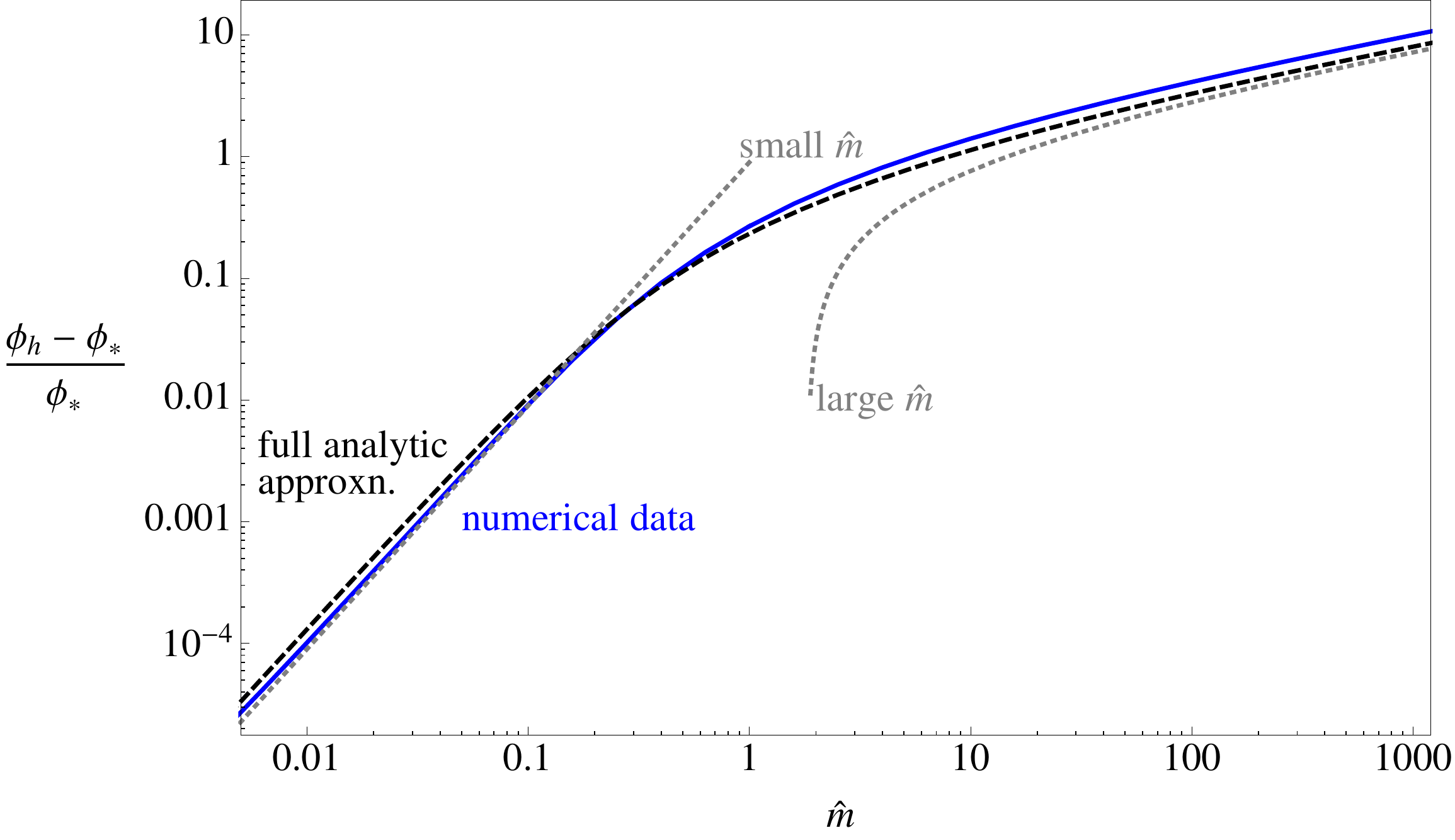}~\nobreak
\includegraphics[scale=0.32]{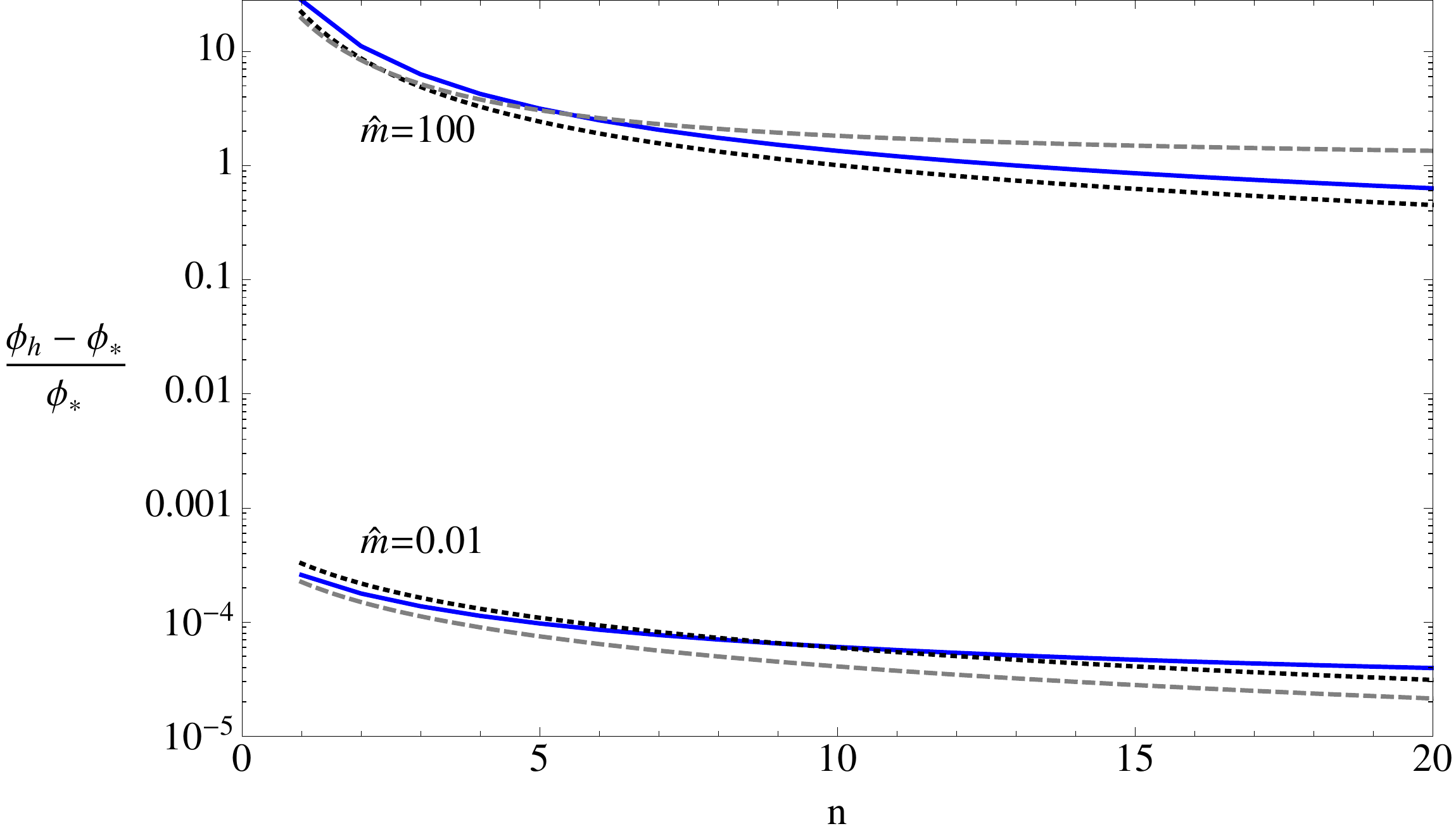}
\caption{The horizon value of the chameleon field as a function of 
${\hat m}$ and $n$, shown for comparison against the analytic 
approximations. The left plot shows the variation with ${\hat m}$, 
and the right the variation with $n$. The numerical data is plotted in blue, 
the full analytic approximation value in dotted black, and the leading
order approximations in dashed grey. In varying $n$, we compare at both 
large and small masses.}
\label{fig:chamhor}
\end{figure}

To confirm the analytic estimates we integrate \eqref{chameq}
numerically for a range of parameters of $\hat m$, $n$, and $x_0$ using
a gradient flow algorithm. In all cases the field profile shows a response 
to the black hole as expected, and we present a selection of our results 
in figures \ref{fig:champlots}--\ref{fig:chamhor} demonstrating 
the qualitative nature of the field profiles, comparing them to the analytic 
approximations, and summarizing the horizon data dependence on 
the model parameters

In figure \ref{fig:champlots} we show the profile of the chameleon field 
over a wide mass range for $R_0=3R_s$ and $n=4$ (the picture is 
similar for different $R_0$ and $n$). 
The numerical solutions show how the profile
of the scalar changes qualitatively between small and large masses. The
low mass scalars remain close to their asymptotic value, and have a fairly
smooth profile. The large mass scalars on the other hand have a much 
sharper fall-off as we approach the dense region, cutting off very strongly
at $R_0$. This is reminiscent of the ``thin shell'' behaviour, \cite{Khoury2004},
which occurs around a massive object. The main difference here is that
it is our exterior region which is dense, with the chameleon relaxing to its
new VEV in the interior. The fact that the field has a significant variation
as it nears $R_0$ is anticipated by our analytic approximation, although is
severely underestimated by the simple geometric approximation used for
lower masses (as seen in comparing the analytic and
numerical solutions in figure \ref{fig:chamcomp}).

In figure \ref{fig:chamcomp} we compare our analytic approximations
to the numerical data. We show sample plots for ${\hat m} = 0.1$ and
${\hat m}=1000$, taking $R_0=3R_s$ and $n=4$ as before. 
In both cases we compare the numerical data to the analytic 
approximation given by \eqref{chregIIapp}, \eqref{chregIapp} with 
the exact expressions from \eqref{chamdata}.
For the large mass plot, we also show what we expect to be the 
more accurate approximation, \eqref{analappchamlg}, derived by
assuming the dominance of the potential. The 
approximations are seen to be extremely accurate in tracking the 
shape of the chameleon profile. The small mass approximation is
accurate to better than 1\%. The large mass plot is accurate near 
$R_0$ (where it was tailored to be), but since $R_0/R_s=3$, the
effect of the geometry rapidly starts to make itself felt, leading to an
overshoot of this approximation at the horizon. However, it can be 
seen to be a far better fit than the geometry dominated expression 
which works so well for small masses.
\FIGURE{
\centering
\includegraphics[scale=0.37]{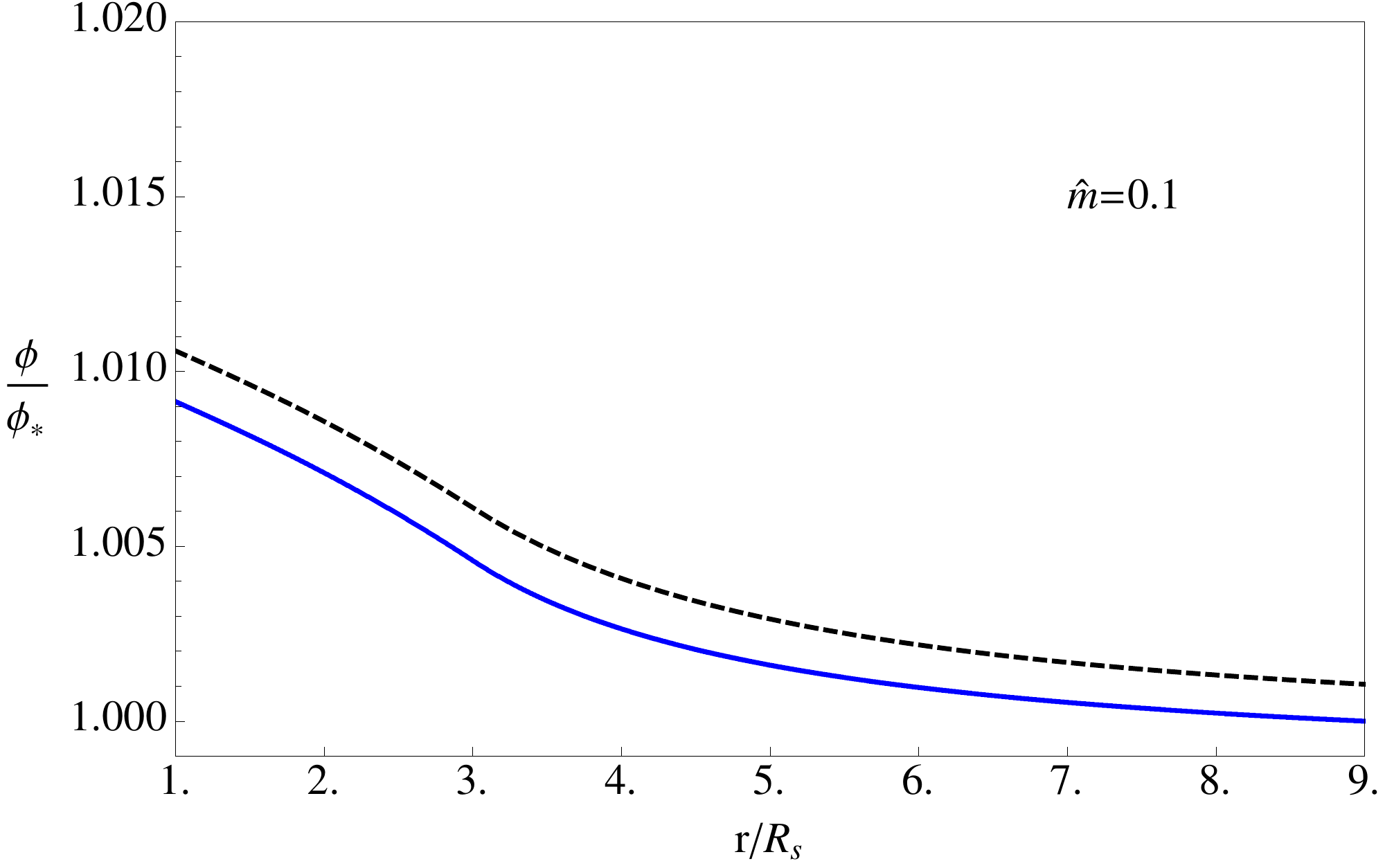}~\nobreak
\includegraphics[scale=0.38]{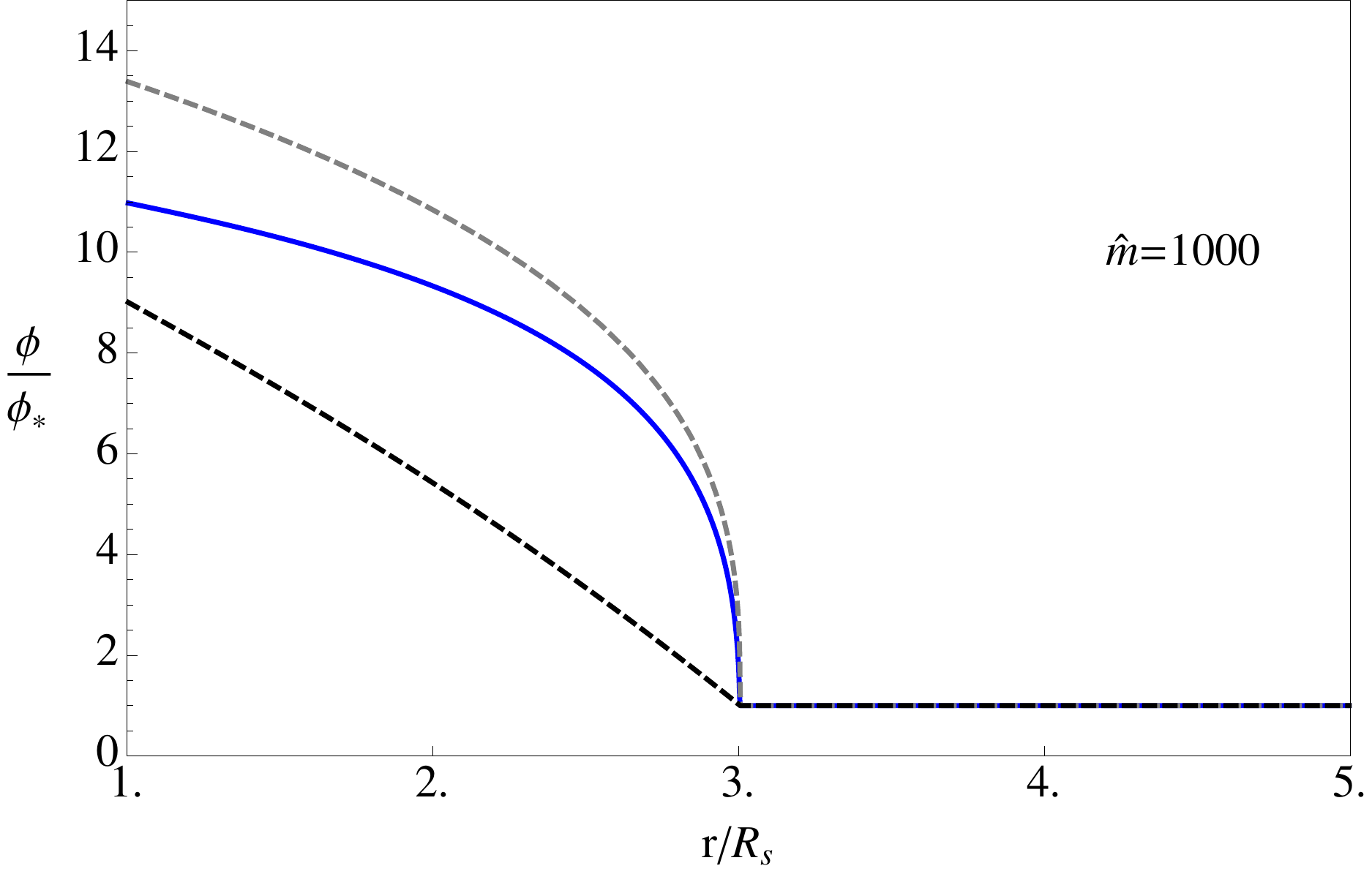}
\caption{
A comparison of the analytic approximations to the numerical
solution (in blue) at both small and large mass. 
The dashed black line is the full analytic
approximation, given by \eqref{chregIIapp}, \eqref{chregIapp} with the 
exact expressions from \eqref{chamdata}. For the large mass case, the
grey line is the potential dominated expression, which better captures the
shape of the chameleon profile.}
\label{fig:chamcomp}
}

Finally, in figure \ref{fig:chamhor} we compare the horizon values of 
the chameleon over a range of masses and the potential index $n$.
We show the numerical horizon data, the analytic data
obtained by solving \eqref{chamdata}, and the small/large mass, large 
$x_0$ leading order approximations to this horizon value. 
In both plots $R_0$ is fixed at $3R_s$, and $n=4$ for varying ${\hat m}$. 
In spite of the shortcomings of the analytic approximations, these plots show 
that the analytics pick up the essential dependence of the chameleon
field. Indeed, it is surprising just how good the rough value is at
extracting the main dependence of the horizon chameleon data
on the model parameters.

\subsection{Dilaton profile}\label{dil_profile}

Environmentally dependent dilatons will also have a non-constant profile.
Recall that the dilaton kinetic term has a coupling factor $k(\phi)$, so 
that the equation of motion for the dilaton is 
\be
\begin{aligned}
\frac{1}{r^2} \left [ r(r-1) \phi' \right]'  &= -
\frac{3\lambda^2\beta\beta'}{1+3\lambda^2\beta^2} \phi^{\prime2}
+ \frac{1}{k^2}
V_{{\rm eff},\phi} (\phi,\rho)
\end{aligned}
\ee
where
\be
\frac{1}{k^2}
V_{{\rm eff},\phi} (\phi,\rho)
=\frac{V_0\lambda^2}{M_p(1+3\lambda^2\beta^2)} \left [ (4\beta -1) 
A^4 e^{-(\phi-\phi_d)/M_p} + \beta A
\frac{\rho}{V_0} \right]
\ee

In region I, $\rho\simeq0$, and hence our vacuum dilaton value is
$\beta=1/4 \Rightarrow \phi_0=\phi_d + M_p/4a_2$, with a mass 
$m_0^2 \simeq 4 a_2V_0/M_p^2$.
In region II, we have instead a small coupling function,
$\beta = V_0/(\rho_\ast+4V_0)$,
and $\phi_\ast \simeq \phi_d + M_p V_0/a_2 \rho_\ast$ with
$m_\ast^2 \simeq b m_0^2$,
where $b=(\phi_0-\phi_d)/(\phi_\ast-\phi_d)\simeq \rho_\ast/4V_0$.
In other words, we have a hierarchy between the two regions set
by the ratio of the local accretion disc or galactic energy density
$\rho_\ast$, and the background potential scale (taken to be the
cosmological density), $V_0$. 
Inputting the parameter values $a_2 \simeq 10^{5}$, 
$V_0\sim \rho_{\text{cosm}}$, we see 
$m_0^2 R_s^2\sim 10^{-40}-10^{-28}$ for astrophysical / supermassive
black holes, hence the scalar field is extremely light in region I. 
In region II, the coupling function is extremely small, hence we expect
$\phi$ to be very close to its minimum $\phi_\ast$. Thus for the dilaton,
we expect a low mass or long range approximation to be appropriate
in any analytical analysis.

Proceeding analogously to the chameleon, we introduce a new field
variable
\be
y = \frac{\phi - \phi_d}{\phi_\ast-\phi_d}\,,
\ee
where we expect $(y-1) \ll 1$, hence $\lambda y \ll b$
(recall $b=\rho_\ast/\rho_{\text{cos}}$),
and our equation of motion is well approximated by
\be
y'' + \frac{2x-1}{x(x-1)} y' = \frac{{\hat m}^2 x}{x-1}
\left [ \Theta[x-x_0] (b-1) y + (y-b) \right ]
\label{dileom}
\ee
where we have set ${\hat m}^2 = \lambda^2 m_0^2 R_s^2$.
Clearly ${\hat m}^2, {\hat m}^2 b \ll 1$, and hence our dilaton 
field will remain close to $y=1$ throughout regions I and II.
We therefore take $y = 1 + \delta y$, and approximate our 
solutions in each region
with the horizon and asymptotic expansions:
\be
y = 1 + 
\begin{cases}
\delta y_h + \frac{{\hat m}^2}{6}(1 + \delta y_h-b) [x^2 + 2x + 2\ln x-3]
& x\leq x_0\\
C\, x^{-(1+{\hat m}\sqrt{b}/2)}\, e^{-{\hat m}\sqrt{b} (x-x_0)} & x \geq x_0
\end{cases}
\label{dilapp}
\ee
where the constants $C$ and $\delta y_h$ are given by matching at $x_0$.
Keeping terms only to leading order in ${\hat m}^2 b$ gives:
\be
\begin{aligned}
\delta y_h &= \frac{{\hat m}^2}{6}(b-1) \,(3x_0^2 + 4x_0 + 2 \log x_0-1)\\
C &= \frac{{\hat m}^2}{3} (b-1)  x_0 (1 + x_0 + x_0^2)\;.
\end{aligned}
\label{dildata}
\ee
We see therefore that the shift in the dilaton value at the horizon
of the black hole is roughly ${\hat m}^2 b x_0^2/2$. 

Numerically, it is rather difficult to explore the extremely small
mass parameter values relevant for the dilaton directly with
our simple techniques, however, by comparing numerical 
data with analytic profiles for a range of less tiny (though still
small) masses, we can verify the analytic understanding, and
extrapolate our results to the mass ranges of relevant for
the dilaton. We therefore integrated \eqref{dileom} numerically
for masses ranging from $10^{-4} - 10^{-14}$, taking again
$R_0=3R_s$.

\begin{figure}
\centering
\includegraphics[scale=0.5]{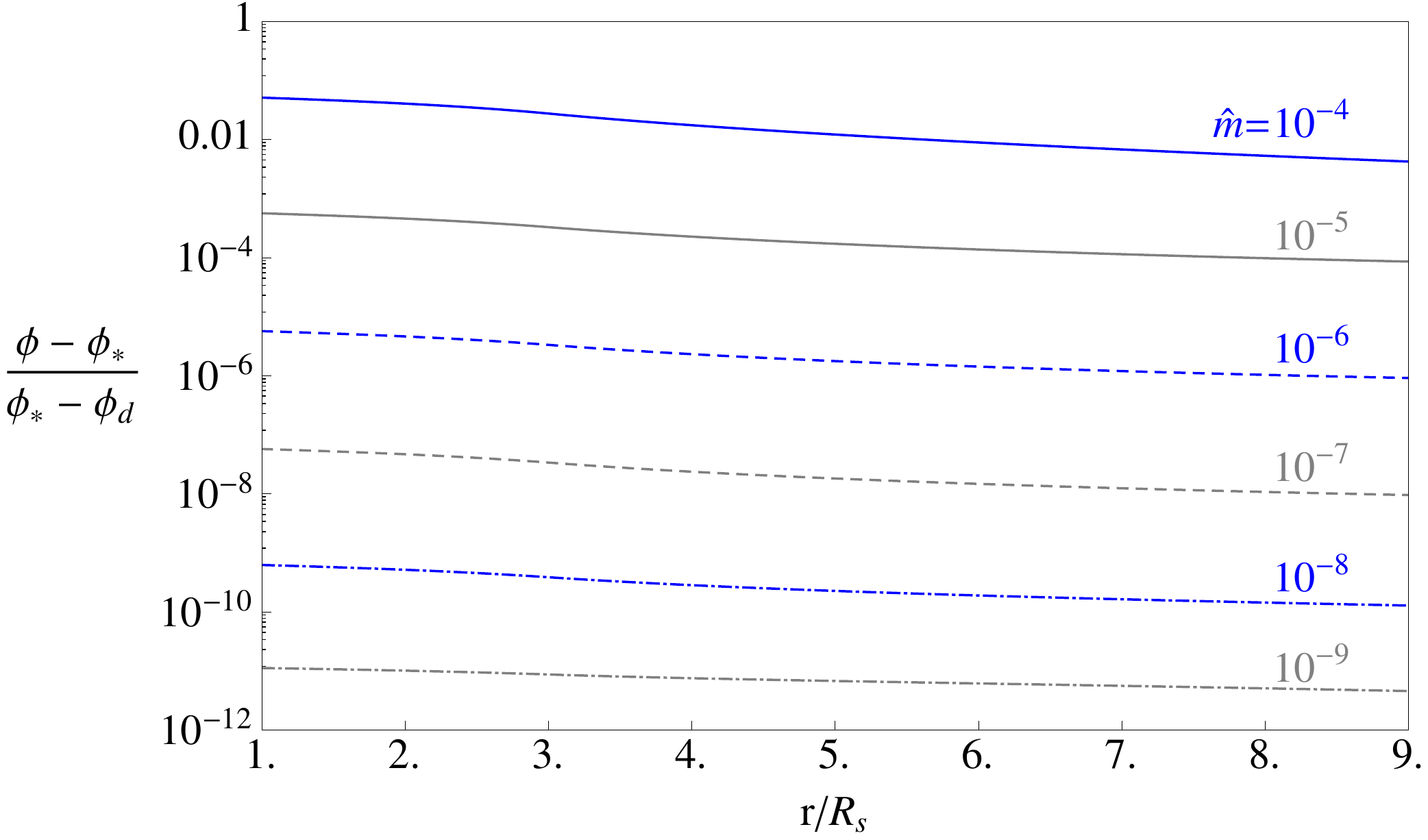}
\caption{A plot of the variation of the dilaton field near the black
hole, shown for a range of masses. The profile of the dilaton
remains similar, with the horizon value dropping as ${\hat m}^2$.}
\label{fig:dilplot}
\end{figure}

In figure \ref{fig:dilplot} we show a plot of the variation of the 
dilaton field over a wide range of orders of magnitude in the mass.
The profile is confirmed to be a gentle, small variation from
the asymptotic value, and drops a couple of orders of magnitude 
for each order of magnitude drop in ${\hat m}$ as expected.

\begin{figure}
\centering
\includegraphics[scale=0.36]{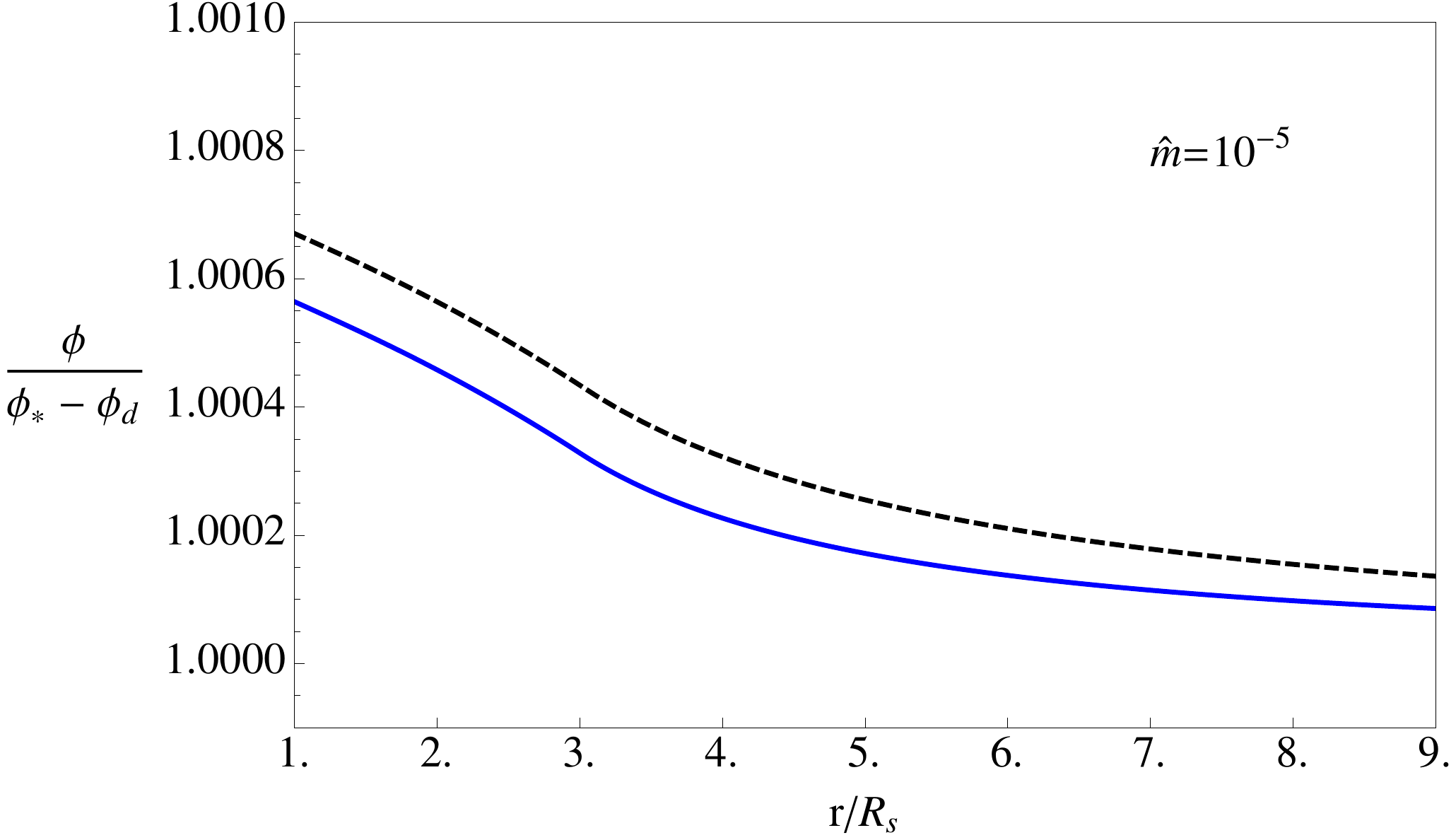}~\nobreak
\includegraphics[scale=0.35]{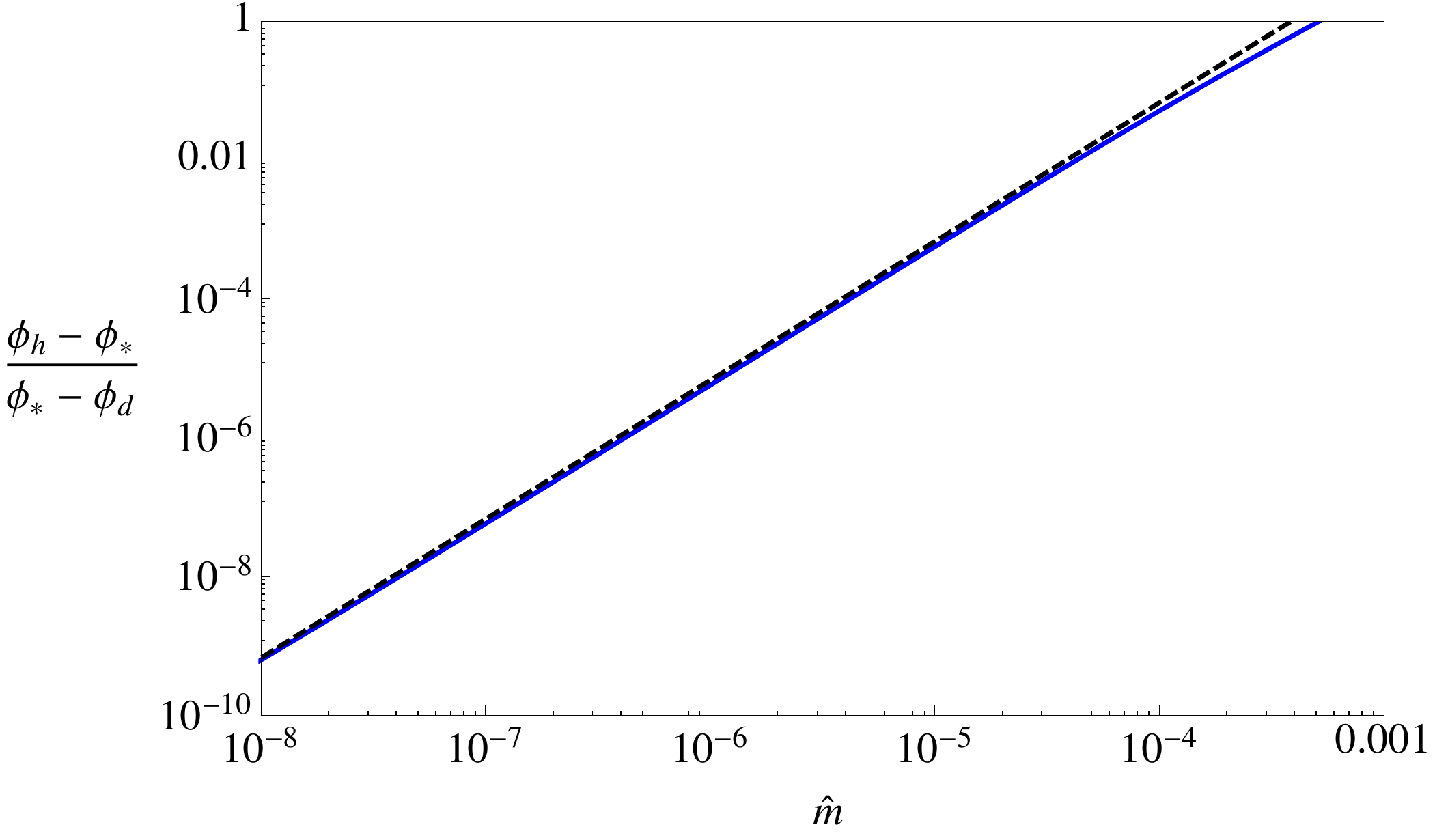}
\caption{A comparison of the analytic approximation to the numerical
solutions: {\it Left:} A plot of the field profile obtained numerically 
for ${\hat m} = 10^{-5}$ compared to the analytic approximation.
{\it Right:} The horizon value of the dilaton field as a function of 
${\hat m}$, shown for comparison against the analytic result.}
\label{fig:dilcf}
\end{figure}

In figure \ref{fig:dilcf}, we compare the numerical and analytic 
solutions as before by plotting an explicit profile (here chosen 
at ${\hat m}=10^{-5}$), and the horizon data as a function of mass, 
again comparing the numerical, and analytical approximations.
As with the chameleon, there is an excellent agreement in the
two expressions, with the horizon data in particular giving almost 
indistinguishable results over a wide range of orders of magnitude 
for ${\hat m}$, therefore we see no reason to doubt the
extrapolation of the data to much smaller masses, or other
model parameters.

\subsection{Symmetron profile}\label{sym_profile}

The symmetron field is distinct from the dilaton and chameleon in its
behaviour, as the screening occurs due to dense regions restoring
symmetry in the coupling function, driving it to zero:
\be
\frac{r - R_s}{r} \phi'' + \frac{2r-R_s}{r^2}\phi'
= \frac{\rho a_2}{M_p^2} \phi - \mu^2 \phi + \lambda \phi^3\,.
\label{symeom}
\ee
In region II, the density $\rhostar$ is large enough so that 
$\langle \phi\rangle =0$ at its minimum, whereas in region I,
$\langle \phi_{\text{vac}}\rangle =\pm \mu/\sqrt{\lambda}$.
Whether or not the symmetron can develop a nontrivial profile
therefore becomes an issue of the tachyonic instability of the 
false vacuum $\langle \phi\rangle =0$ in region I, which is always
a solution to the symmetron equation of motion \eqref{symeom}. 
In order for $\phi$ to develop a nonzero profile, there has to be
sufficient space for a fundamental mode of the wave equation
to exist within region I, roughly of order the Compton wavelength
of the symmetron. Given that we take $R_0=3R_s$, this
translates to $\mu R_s \gtrsim {\cal O} (1/3)$ or so. Given that
we expect $\mu^{-1} \sim $Mpc, we are clearly well outside
this r\'egime for any cosmologically relevant symmetron, 
therefore our analysis will focus on confirming this intuition,
and demonstrating the existence and magnitude of this limit.

Writing $x=r/R_s$ as usual, ${\hat \phi} = \sqrt{\lambda}
\phi/\eta$, ${\hat\mu}=\mu R_s$, and ${\hat m}^2 = \rho_\ast a_2
R_s^2/M_p^2 - {\hat\mu}^2$, the equation of motion becomes:
\be
\left [ x(x-1) {\hat\phi}' \right]' = x^2 {\hat\phi}
\left [  {\hat m}^2 \Theta(x-x_0) -{\hat\mu}^2\Theta(x_0-x) 
(1-{\hat \phi}^2)\right]\,.
\ee
For small mass parameters, we might expect a small variation in our 
scalar field, therefore we try our usual approximate solution for a slowly
varying field,
\be
{\hat\phi} = \begin{cases}
{\hat\phi}_h - \frac{{\hat\mu}^2}{6} {\hat\phi}_h (1-{\hat\phi}_h ^2)
\left [ x^2+2x+2\log x -3 \right] & x\leq x_0\\
C\,\frac{e^{-{\hat m}(x-x_0)}}{ x^{-(1+{\hat m}/2)}} & x \geq x_0\,.
\end{cases}
\label{symapp}
\ee
However we now see something interesting arising 
in our matching conditions:
\be
\begin{aligned}
\frac{C}{x_0^{1+{\hat m}/2}} &=
{\hat\phi}_h \left [ 1 - \frac{{\hat\mu}^2}{6} (1-{\hat\phi}_h^2)
(x_0^2+2x_0 +2\log x_0 -3) \right] \\
&=
\frac{{\hat\mu}^2}{3} {\hat\phi}_h (1-{\hat\phi}_h^2)
\frac{(x_0^2+x_0 +1)}{({\hat m} x_0 + 1 + {\hat m}/2)}
\end{aligned}
\ee
${\hat\phi}_h$ now scales out of these relations, and we see that 
$(1-{\hat\phi}_h^2) \simeq \frac{ 6}{{\hat\mu}^2 x_0^2}\leq 1$.
Clearly this is inconsistent for small $\hat\mu$, therefore our approximation
indicates that there should be a lower bound on the mass for which
a symmetron solution can exist (in keeping with our earlier intuition),
and gives the mass limit as ${\hat\mu}^2 x_0^2 \lesssim 6$. 

For large masses, where we would expect ${\hat\phi}$ to
approach close to unity very rapidly inside region I, the approximation we
use to get the interior solution in \eqref{symapp} is not reliable, as
${\hat\phi}$ is varying significantly. With the chameleon, 
we approximated our profile by taking the potential to dominate the
behaviour of the scalar, for the symmetron, provided $x_0\hat \mu$ 
is large enough, our differential equation is well approximated by the
$\lambda \phi^4$ kink model -- and our field will take on a tanh profile
as it makes the transition to its new, true, vacuum.
We can therefore take the approximation
\be
{\hat\phi} \simeq \begin{cases}
\text{tanh} \left ( \frac{{\hat\mu}(x_0+ \delta x-x)}{\sqrt{2}}\right) & x<x_0\\
\tanh \left ( \frac{{\hat\mu} \, \delta x}{\sqrt{2}}\right)
\left ( \frac{x_0}{x} \right ) ^{1+{\hat m}/2} e^{-{\hat m}(x-x_0)} & x> x_0
\end{cases}
\ee
which rapidly transitions from $0$ to $1$ in a thin shell inside region I.

Obviously, these arguments are only suggestive, and in no way constitute
a proof of the nonexistence of a nontrivial solution at low mass, however,
they are consistent with our numerical findings (see appendix \ref{appsymm}), 
and also with the results of
\cite{Cardoso2013}. Indeed, studying our black hole model in the context 
of the analysis of spontaneous scalarization in \cite{Cardoso2013}, we 
again find that at low masses the black hole is not required 
to have a nontrivial symmetron profile.

\section{Observational Implications}\label{implications}

In the previous section we studied static, spherically symmetric solutions 
of the scalar field equation on a background comprising a Schwarzschild 
black hole and a static, spherically symmetric, constant matter distribution. 
Having verified that the analytic estimates are a good approximation 
to the full solution we will now use them to compute astrophysical effects.   

In our static model, the black hole is not moving through the scalar 
gradient and  $\phistar$ is constant, so on its own it will not emit scalar 
radiation through the mechanisms explored in~\cite{Horbatsch2011} 
and~\cite{Berti2013}. However, as the presence of a non-zero scalar 
gradient will cause matter particles to feel a fifth force, we 
can comment on how chameleon and dilaton hair will affect accretion 
onto the black hole. Though it is possible that the fifth force effects 
could alter the structure of  accretion disks~\cite{Perez2013}, determining 
whether the effect could be observable would require astrophysical 
modelling beyond the scope of this paper. We can, 
however, study its effects on the dynamics of an orbiting test particle.

For a test particle at distance $r$ from the black hole, the ratio of the fifth 
force to that of Newtonian gravity is 
\be\label{forcecomp}
\frac{|F_{\phi}|}{|F_N|}\approx \left(\frac{r}{R_s}\right)^2 
\beta(\phi)|\vec{\nabla}\phi|\frac{M_{BH}}{M_p^3}.
\ee
Our initial assumptions guarantee this  to be small:  In \textsection\ref{profile}
we stated that a requirement for the Schwarzschild metric to be a good 
leading order background was  
$|T_{\phi}|\sim |\nabla\phi|^2\ll \frac{M_p^6}{M_{BH}^2}$.  
For $\beta\sim\mathcal{O}(1)$, this condition is identical to requiring 
$|F_{\phi}/F_N|\ll 1$.  For the profiles presented in \textsection\ref{profile}, 
the largest fifth force will be generated by chameleons which vary rapidly 
in a thin shell near $R_0$. Inserting even the most optimistic 
parameter values shows that the ratio in \eqref{forcecomp} will be 
at most $\mathcal{O}(10^{-2})$. 
To evaluate the relevance of our estimated scalar gradients, we should study 
them in comparison to another small effect, namely gravitational radiation.

We can do this if we view our static black hole model  as the supermassive 
partner of an extreme mass ratio inspiral (EMRI) binary system. Such EMRI 
systems, which consist of a stellar mass compact object orbiting a supermassive 
black hole, will be detectable by future space-based gravitational wave 
detectors~\cite{Yunes2012}.   In GR they will emit gravitational radiation 
at a rate approximated to leading order in $\dot r$ by,
\cite{Will:1993ns,Peters1963},
\be\label{quadrad}
\frac{dE}{dt} = -\left\langle \frac{m_t^2G^3M_{BH}^2}{c^5 r^4}
\frac{8}{15}(12 v^2 - 11\dot{r}^2)\right\rangle,
\ee
where  $m_t$ is the mass of in-falling object, $v$ its velocity, $r$ 
its radial position, and the angled brackets indicate an average over 
an orbital  period.  This emission will cause the orbiting object to 
gradually move to smaller radii, which we would detect as a decrease 
in the period of detected gravitational waves.

To leading, Newtonian, order, the gravitational binding energy 
of the system, per unit mass of the test particle, will be
\be
\mathcal{E} = \frac{G M_{BH}}{r}- \frac{ v^2}{2}.
\ee
The  conservation of total energy allows us to relate the rate of change of this 
Newtonian energy to flux being carried away by gravitational waves 
$\dot{\mathcal{E}}_{GR}$ and interactions with scalar fifth forces 
$\dot{\mathcal{E}}_{\phi}$.
\be
\dot{\mathcal{E}} = \dot{\mathcal{E}}_{GR} + \dot{\mathcal{E}}_{\phi}.
\ee
The evolving frequency of gravitational waves from a binary system 
tells us about rate of change of its orbital period, allowing us to 
measure $\dot{\mathcal{E}}$. Therefore, if the scalar fifth force 
can generate effects comparable to those of gravitational waves, it 
could potentially be detected by observations of EMRI systems.

\subsection{Classical energy estimate}

A detailed examination of the chameleon, dilaton, and symmetron effects on 
EMRI signals would require a sophisticated treatment like that in
\cite{Yunes2012}, taking into account time-dependent effects on the scalar field 
as well as the fact that an accretion disk will affect EMRI dynamics even in GR
\cite{Kocsis2011,Yunes2011a}.  However, we find that if we model the stellar 
mass object as a test particle in orbit around the black hole, we are able to 
present an order-of-magnitude estimate for how the presence of a scalar 
fifth force will affect its gravitational binding energy.

The classical equation of motion for a test particle with mass $m_t$ orbiting 
a black hole  in the presence of a radial fifth force of magnitude $m_ta_{\phi}$ 
can be manipulated to show 
\be
\dot{\mathcal{E}}_{\phi} = v a_{\phi}.
\ee
Fifth force effects on in-falling particles are most likely to be important 
near $r=R_0$. Here we can estimate $|\dot{\bf{r}}|$ for an object in a 
bound orbit by equating centripetal acceleration with that of gravity, 
giving $v=  \sqrt{\frac{GM}{R_0}} = \sqrt{\frac{R_S}{2R_0}}.$  
We note that at the innermost stable orbit $R_0=3 R_S$, $v\sim 
\frac{1}{\sqrt{6}}\sim 0.4$.  Such a large velocities signify a breakdown of 
Newtonian mechanics, however, since our goal is to derive an order of 
magnitude estimate of the effect, the approximation suffices.

We can now use our results from \textsection\ref{profile} to make 
a rough estimate for the scalar gradient, and thus, for the magnitude 
of the fifth force.  When the scalar field mass in Region II is large, such
as the case for the chameleon at accretion disc densities,
we found that the field varies rapidly in a thin shell at the 
edge of the matter distribution. We can use our analytic approximation
to estimate the scalar gradient near $R_0$ as 
\be
\frac{d\phi}{dr} \bigg |_{R_0} \sim (\phi_\ast - \phi_h)
\frac{m_\ast}{\left (m_\ast R_0\right)^{2/(n+2)}}\,.
\ee
For chameleons at galactic densities, and for dilatons, $\mstar R_0\ll 1$, 
so the scalar will vary slowly everywhere. In this case we 
should instead write $\phi,_r\approx(\phistar - \phi_h)/R_0 $.

Using these results we can estimate that when a test particle is 
near $R_0$, its rate of change in energy per unit mass due to 
the fifth force will be
\be
\dot{\mathcal{E}}_{\phi} \approx \left(\sqrt{\frac{R_S}{2 R_0}}\right)
\frac{\beta}{M_p}\left (\frac{\phi_h- \phistar}{\Delta R}\right),
\ee
where $\Delta R$ is the smaller of $R_0$ or $R_0^{\frac{2}{n+2}}
\mstar^{-\frac{n}{n+2}}$.

We note that for chameleons and dilatons $\phi_h>\phistar$, so 
$\dot{\mathcal{E}}_{\phi}$ will be positive.  This is a reflection of the fact 
that the scalar force will manifest itself as an extra attraction between the 
test particle and the matter distribution $\rho(r)$.  This can also be seen if 
we assume the test particle is a stellar mass black hole and write its scalar 
`charge', in the sense of \cite{Horbatsch2011,Berti2013,Jacobson1999}, 
$Q(t)\propto \dot{\bf{r}}(t)\cdot\bf{\nabla}\phi(\bf{r}(t))$.  
We note that as the stellar mass black hole moves inwards through the 
scalar gradient, its `charge' will increase, indicating that it will absorb energy 
from the scalar field.  This suggests that the scalar profile generated by 
the interaction of screened modified gravity and an accretion disk 
or galactic halo will slow the inspiral of an EMRI system.  

\subsection{Comparison to quadrupole radiation in GR}

Comparing the scalar radiation to the gravitational wave effects gives
\be\label{energycomp}
\left|\frac{\dot{\mathcal{E}}_{\phi}}{\dot{\mathcal{E}}_{GR} }\right| 
\sim \beta(\phistar)\left(\frac{R_0}{R_s}\right)^{\frac{9}{2}}
\left(\frac{\phi_h- \phistar}{\Delta R}\right)\frac
{M_{BH}}{M_p^3}\left[\frac{M_{BH}}{m_t}\right].
\ee
The first part of this expression is the same (small) term that 
appeared in \eqref{forcecomp}, indicating that the scalar fifth force 
will be small compared to that of Newtonian gravity. Because we are 
considering an EMRI system where $m_t\ll M_{BH}$, the final
bracketed term will be large.  Therefore, to be able to evaluate whether 
the scalar field induced energy evolution is comparable to that from 
quadrupole radiation in GR, we must insert physically relevant values 
for the various parameters.

Let us assume that the test mass is a solar mass black hole being 
captured by a supermassive black hole, which we will take to have
a mass either $M_{BH}\sim 10^{6-9} M_{\odot} \sim 10^{44-47}M_p$.   
We can now use our results from \textsection\ref{profile} to estimate 
this relation for the scenarios where we predict a non-zero scalar gradient.\\

\noindent\textbf{Short range chameleons}\\

Chameleons are expected to have $\mstar R_0\gg 1$ when the 
central black hole is surrounded by an accretion disk.   
For this case we set $\Delta R= R_0^{\frac{2}{n+2}}
\mstar^{-\frac{n}{n+2}}$, and substituting for 
$\phi_h$ (eq.\ \eqref{ch_phih}) gives 
\be
\left|\frac{\dot{\mathcal{E}}_{\phi}}{\dot{\mathcal{E}}_{GR} }\right| 
\sim 10^{-28-60/(n+2)} \, \beta_\ast \frac{M_{BH}^2}{M_pm_t}
\left(\frac{m_\ast}{M_p}\right)^{\frac{n}{n+2}}
\ee
where  we have set $M\sim 10^{-3}\,\text{eV}$. 
If we assume that $\beta(\phistar)\sim\mathcal{O}(1)$ 
and reference Table~\ref{ch_masstable} for chameleon masses, we find 
\be
\left|\frac{\dot{\mathcal{E}}_{\phi}}{\dot{\mathcal{E}}_{GR} }\right| 
\sim 10^{-23+ \frac{2(n+3)}{(n+1)(n+2)}} \,
\left ( \frac{M_{BH}}{M_\odot}\right )^2
\approx 10^{-11} - 10^{-5}
\ee
for the super (or super-super) massive black holes,
relatively independent of the index, $n$. 
Thus, chameleon fifth force effects on the test particle's dynamics 
will be much smaller than those of gravitational radiation.  
We note that because the non-zero fifth force operates only in a thin shell 
very near the ISCO, the effects will likely be more 
suppressed than this estimate suggests. However, should
the chameleon model have a significant coupling parameter
$\beta$, then these conclusions will be changed, as the ratio
is proportional to $\beta$.\\

\noindent\textbf{Long range chameleons}\\

When the black hole is surrounded by matter with galactic halo density, 
it is possible that the chameleon will be light compared to the size of 
the system.  For this case we use $\Delta R = R_0 \simeq 3R_S$ to get
\be \label{light_ecomp}
\left|\frac{\dot{\mathcal{E}}_{\phi}}{\dot{\mathcal{E}}_{GR} }\right| 
\sim \beta \frac{(\phi_h - \phistar)}{M_p}\frac{M_{BH}}{m_t}.
\ee
The chameleon profile calculations in Section~\ref{ch_profile} 
then give
\be
\left|\frac{\dot{\mathcal{E}}_{\phi}}{\dot{\mathcal{E}}_{GR} }\right| 
\sim 10^{2} \beta^2 \, \frac{\rhostar R_0^2}{2M_p^2}
\frac{M_{BH}}{M_\odot}
\sim 10^{-42} \beta^2 \, \frac{\rhostar}{\rho_{\text{cos}}}
\left ( \frac{M_{BH}}{M_\odot} \right ) ^3 \sim 
10^{-18} - 10^{-9}
\ee
where we assume $\beta\sim \mathcal{O}(1)$ and 
$\rho_\ast \sim 10^6\rho_{\text{cos}}$. Once again, if the coupling
function becomes appreciably large, this conclusion will change.\\

\noindent\textbf{Dilatons}\\

Dilatons will always always satisfy $\mstar R_0\ll 1$, so we again 
use \eqref{light_ecomp}.  From section \ref{dil_profile}, we find
\be
\frac{\varphi_h - \varphistar }{M_p} \approx 
k(\phi_\ast)\delta\phi \sim \lambda \beta_\ast
\frac{\rho_\ast R_0^2}{M_p^2} =\lambda 
\frac{\rho_{\text{cos}} R_0^2}{M_p^2} 
\approx 10^{-42}  \lambda 
\left (\frac{M_{BH}}{M_\odot}\right)^2
\ee
using $\beta_\ast \simeq \rho_{\text{cos}}/\rho_\ast$. We therefore
find 
\be
\left|\frac{\dot{\mathcal{E}}_{\phi}}{\dot{\mathcal{E}}_{GR} }\right| 
\sim 10^{-42}\, \lambda \frac{\rho_{\text{cos}}}{\rho_\ast}
\left (\frac{M_{BH}}{M_\odot}\right)^3
\sim [10^{-24}- 10^{-15} ] \lambda \left(\frac
{\rho_{\text{cosm}}}{\rhostar}\right).
\ee
Clearly dilatons have a far weaker impact than chameleons on 
radiative loss.

\section{Conclusions}\label{conclusion}

The strong field, large curvature limit of modified gravity is a largely 
unexplored area for constraining various model dependent parameters 
and differentiating between the numerous models that exist in literature. 
We have presented an exploratory calculation for theories of modified 
gravity with screening mechanisms by studying a static, spherically 
symmetric black hole with an $r$-dependent matter distribution around it. 
In this construction, we found that chameleon and dilaton fields develop 
a non-trivial scalar profile, while the symmetron assumes a 
constant value. An order of magnitude estimate showed that the 
resulting scalar gradients affect in falling test particles in a way that 
is sub-leading compared to the quadruple radiation in GR and thus 
would be observationally challenging to detect.   

Note that our findings for symmetrons could be qualitatively 
altered if we used a different matter distribution in our setup.  
As we discussed in \textsection\ref{sym_profile}, 
if the matter density far from the black hole is small enough to make the 
symmetron approach $\phistar=\mu/\sqrt{\lambda}$ as its boundary 
condition, its field equation will no longer have a constant solution.   
This could occur, for example, if we used a model of an accretion disk 
which had finite spatial extent.  We would, however, expect any 
symmetron gradients to be comparable to the ones found above for long 
range chameleons and dilatons. 

Similar reasoning can be applied when we consider solutions to 
the scalar field equation with angular dependence, or solutions on a 
Kerr background.   No hair theorems  require that in a stationary 
system involving matter around a black hole, any gradient in 
the scalar field will be sourced entirely by non-uniformities in matter 
density.  If we assume that the density contrasts examined above are 
typical of astrophysical matter distributions near black holes, the 
magnitude of the scalar gradients produced (and their consequent 
fifth forces) should likewise be representative.  Thus generically we 
would expect theories of screened modified gravity to produce solutions 
distinct from GR around stationary black hole systems with sub leading 
corrections to various observational effects.

It would be interesting to explore this strong field limit further. One 
possible physical effect to study within our setup would be the phase 
of gravitational waves. Our crude estimates suggest that the rate of 
GR quadruple radiation will dominate over any radiation in the scalar 
sector, however,  in~\cite{Berti2012}, the authors showed that in 
Brans-Dicke theory the presence of a massive scalar field affects 
the phasing of gravitational radiation from binary systems significantly 
and they found that for observations of  intermediate mass ratio inspirals, 
this effect could be used to place constraints on model parameters (a 
lower bound on $\omega_{BD}$ and an upper bound on the mass 
of the scalar) which are competitive with Cassini and LLR measurements.

Additionally, the requirement that our system and solutions be stationary 
causes us to disregard the possibility of transient effects associated 
with superradiance.  Superradiance is a property of rotating black holes 
through which incident waves with at resonant frequency can become 
amplified by extracting some of the black hole's rotational energy.  
A number of studies have shown that  superradiant effects can give 
rise to long-lived unstable modes in the presence of a massive scalar
\cite{Cardoso2005,Dolan2007,Dolan2012,Witek2013,Cardoso2013a}.  
This occurs for isolated black holes, but it also has been shown that 
when a black hole is surrounded by matter, superradiant modes can be 
amplified by factors as large as $10^5$~\cite{Cardoso2013}.   
If the scalar's mass is very light, the instability timescale for these 
modes becomes short and would result in gaps in the mass-spin 
phase space of observed black holes.   
Because of this, measurements of a black hole's mass and spin 
can be used to constrain allowed masses for scalar fields.  This 
technique has already been used to place the most stringent upper 
bound on the mass of the photon~\cite{Cardoso2013a,Berti2013a}.

Yet another potentially observable effect of superradiance occurs 
in EMRI systems, in what is known as floating orbits
\cite{Yunes2013,Yunes2012,Cardoso2011}.  Floating 
orbits occur when the orbital frequency of the small compact object  
can excite superradiant modes.  This causes the scalar field to 
transfer rotational energy from the central black hole into the 
orbit of the small compact object, counteracting the energy 
lost through quadrupole radiation and slowing the inspiral.   
Such floating orbits would affect the EMRI's gravitational waveforms 
significantly, so the detection of an EMRI signal consistent with GR 
would allow us to place strong constraints on the mass of a light scalar.  

We note with interest that these effects are typically studied in the 
context of scalar fields with a mass in the range  
$10^{-33}\,\text{eV} - 10^{-10}\,\text{eV}$~\cite{Berti2013a}, 
which is relevant for dilatons and symmetrons, as well as chameleons at 
galactic or cosmological densities.  Consideration of gravitational 
wave phasing, superradiant instabilities, and floating orbits will 
thus be important if we want to understand how these theories 
can be constrained by observations of black hole systems. 
We hope to further explore some of these effects in future works. 

\acknowledgments

RG is supported in part by STFC (Consolidated Grant ST/J000426/1),
in part by the Wolfson Foundation and Royal Society, and in part
by Perimeter Institute for Theoretical Physics. 
Research at Perimeter Institute is supported by the Government of
Canada through Industry Canada and by the Province of Ontario through the
Ministry of Research and Innovation. RJ is supported by the 
Cambridge Commonwealth Trust and Trinity College, Cambridge. 
ACD is supported in part by STFC. 
JM was supported in Cambridge by a Marshall Scholarship.
 
This work was also supported in part by the National Science Foundation 
under Grant No. PHYS-1066293 and the hospitality of the 
Aspen Center for Physics.

\appendix
\section{Numerical results for the symmetron}
\label{appsymm}

As previously noted, the mass of the symmetron is well below the 
expected threshold for a nontrivial field profile, therefore our numerical 
investigation focussed on the confirmation of the analytical picture 
presented above. We took  $R_0 = 3R_s$ as usual, and explored a 
range of masses around unity, ${\hat\mu}^2 \sim 0.5-5$, 
to cover the r\'egime in which we expect the symmetron
field switches on in the vicinity of the black hole.

\begin{figure}
\centering
\includegraphics[scale=0.50]{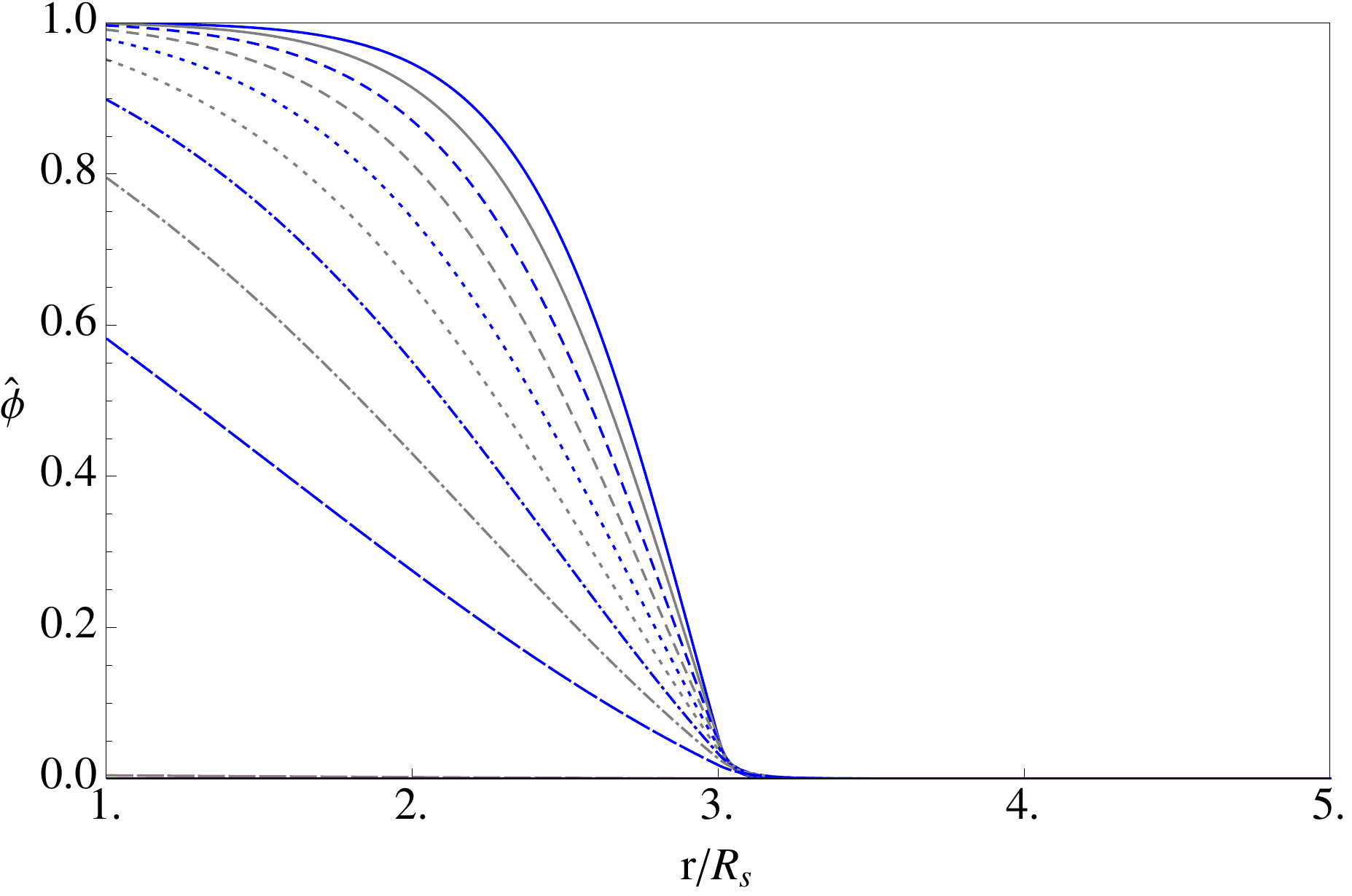}
\caption{
The symmetron profile for a sequence of masses from
$\sqrt{5}$ to $1/\sqrt{2}$:
${\hat\mu}^2=5\times 10^{-i/10}$ for $i=0,1,....10$. 
This range covers the switching off of the
symmetron in region I, and the final profile can just be detected on the 
$x-$axis.}
\label{fig:symmplots}
\end{figure}

\begin{figure}
\centering
\includegraphics[scale=0.37]{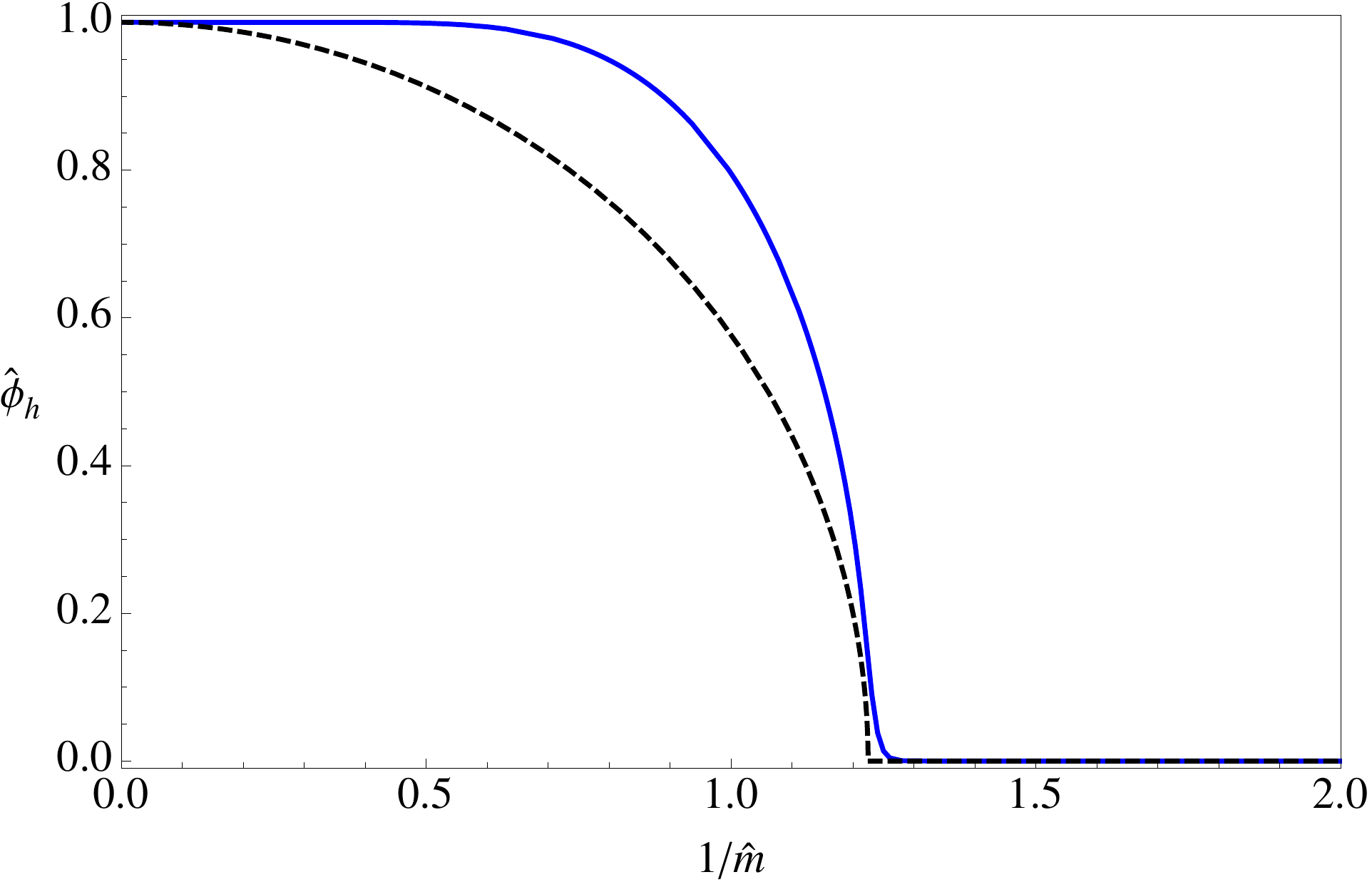}\nobreak
\includegraphics[scale=0.37]{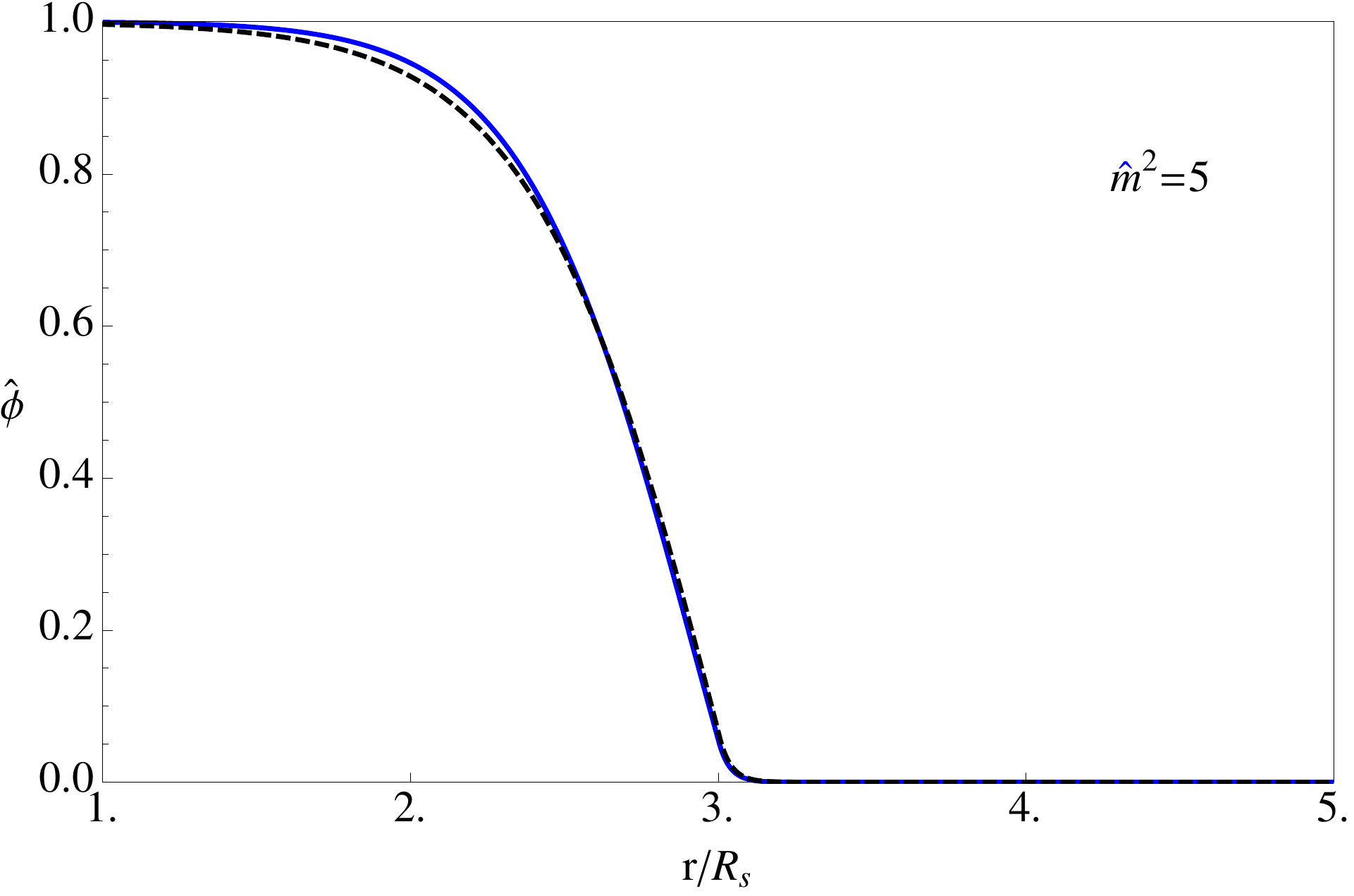}
\caption{Comparing the analytics and numerics for the symmetron.
On left the computed horizon value of the symmetron is shown in blue,
together with the approximation in dashed black. On the right, the
symmetron profile for ${\hat\mu}=\sqrt{5}$, ${\hat m}=10{\hat\mu}$ 
is shown (in blue) compared to
the analytic tanh profile (dashed black) approximation. Even though 
the mass is not particularly large, the approximation is extremely good.}
\label{fig:symmhor}
\end{figure}

The results presented
in figures \ref{fig:symmplots} and \ref{fig:symmhor} were computed
for a galactic density environment, $\rho/V_0 \sim 10^6$, however
the system is insensitive to an increase in density of region II, as the
key physics here is that the coupling function to matter has switched 
off, fixing the symmetron at its (local) vacuum value.

Figure \ref{fig:symmplots} shows the profile of the symmetron gradually 
switching on as the effective mass in the vacuum of region I is
raised. The transition of the horizon value of the symmetron 
from zero to unity as the mass is increased is
seen to occur rather rapidly, and is shown in more detail in figure
\ref{fig:symmhor}. The key point from figure \ref{fig:symmplots} 
is that the intuition of a cut-off in mass for a nontrivial symmetron
profile is confirmed. 

In figure \ref{fig:symmhor}, we explore the correspondence between
the analytic analysis and the numerical data. First, the 
phase transition of the horizon value of the symmetron as a function
of inverse mass is shown in comparison to the low mass
analytic approximation, $\sqrt{1-6/{\hat\mu}^2 x_0^2}$. The
cut-off in the symmetron profile is seen to be quite sharp, and
unsurprisingly the low mass analytic approximation does not track 
the initial drop in the horizon value at larger masses that well, although
it is very accurate at predicting the switch-off of the symmetron.
Secondly, the field profile of a larger mass (${\hat\mu}=\sqrt{5}$) 
symmetron is shown for comparison against the tanh profile of the 
analytic guess.
Here, the correspondence between the analytic approximation and 
the actual profile is amazingly good and shows the analytic
work is capturing the essence of the actual physics very well.


\providecommand{\href}[2]{#2}\begingroup\raggedright

\endgroup


\end{document}